\documentclass[12pt,preprint]{aastex}

\newcommand{\vecrho}{\vec{\rho}}
\newcommand{\amt}{\alpha_m(t)}

\newcommand{\vobs}{\{V_{mn}^{obs}(u_k,v_k,t_k),\ k=1,N_{mn},\ \forall\ (m,n)\}}
\newcommand{\vmn}[1]{V_{mn}^{#1}(u_k,v_k,t_k)}
\newcommand{\vmod}{\left (\sum_{l=0}^{N_p} a_l t_k^l + j \sum_{l=0}^{N_p} b_l t_k^l \right )}
\newcommand{\ip}{S(\vecrho)}
\newcommand{\hip}{\hat{S}(\vecrho)}
\newcommand{\gma}{G_m^{\kappa}}
\newcommand{\gnac}{G_n^{*\kappa}}
\newcommand{\hgma}{\hat{G}_m^{\kappa}}
\newcommand{\tma}{T_m^{\kappa}}
\newcommand{\htma}{\hat{T}_m^{\kappa}}
\newcommand{\mv}{\mathcal{V}}
\newcommand{\fip}{\mathcal{F}_{\hip}}
\newcommand{\hfip}{\hat{\mathcal{F}}_{\hip}}
\newcommand{\pv}{p(\mathcal{V};\ip,\gma,\tma)}
\newcommand{\tpv}{\tilde{p}(\mathcal{V};\ip,\gma,\tma)}
\newcommand{\dtb}{\triangle t_b}
\newcommand{\ntt}{N_{\triangle t}}
\newcommand{\nant}{N_{ant}}
\newcommand{\nt}{N_t}

\slugcomment{To appear in AJ}

\shorttitle{Bootstrap resampling for fidelity assessment}
\shortauthors{Kemball \& Martinsek}

\begin{document}
\title{Bootstrap resampling as a tool for radio-interferometric \\
       imaging fidelity assessment}

\author{Athol Kemball}
\affil{National Center for Supercomputing Applications, University of
    Illinois at Urbana-Champaign, 605 E. Springfield Avenue,
    Champaign, IL 61820}

\and

\author{Adam Martinsek}
\affil{Department of Statistics, University of Illinois at
Urbana-Champaign, 725 S. Wright Street, Champaign, IL 61820}

\begin{abstract}

We report on a numerical evaluation of the statistical bootstrap as a
technique for radio-interferometric imaging fidelity assessment. The
development of a fidelity assessment technique is an important scientific
prerequisite for automated pipeline reduction of data from modern
radio interferometers. We evaluate the statistical performance of two
bootstrap methods, the model-based bootstrap and subsample bootstrap,
against a Monte Carlo parametric simulation, using interferometric
polarization calibration and imaging as the representative problem
under study. We find both statistical resampling techniques to be
viable approaches to radio-interferometric imaging fidelity assessment
which merit further investigation. We also report on the development
and implementation of a new self-calibration algorithm for
radio-interferometric polarimetry which makes no approximations for
the polarization source model.

\end{abstract}


\section{Introduction}

Radio interferometry produces astronomical images of unmatched
spatial resolution and of unique scientific value. The images are
formed by solving an inverse imaging problem connecting the spatial
coherence of the incident radiation, gridded into the visibility plane
orthogonal to the source direction, and the radio brightness
distribution projected on the plane of the sky. An extensive review of
this observing method is provided by \citet{tho01}. The techniques
used for indirect imaging in this discipline are well-established and
robust, when used within the realm in which they are mathematically
applicable. Best practices have been established in the community for
their practical application based on algorithm research and
evaluation, and common data reduction experience \citep{per89,tay99}.

A central unsolved question, however, concerns the development of a
quantitative method to assess the fidelity and precision of the
reconstructed images at each pixel on the plane of the sky. Radio
synthesis image formation requires a simultaneous solution for the
source brightness distribution and instrumental calibration effects,
including the effects of atmospheric propagation and the instrumental
response of the antennas and receiving electronics of the system
\citep{rog74}. This joint solution is generally solved iteratively
\citep{rea78}, with an image deconvolution step per iteration to
remove the effects of the sparse sampling of the measured correlation
data in the visibility plane \citep{hog74}. Errors in calibration and
imaging both introduce errors into the final reconstructed
images. Formally, the inverse imaging problem admits an infinite
family of solutions for the source brightness distribution due to the
sparse sampling of the visibility data. However, the problem can be
regularized by applying physically reasonable constraints, such as
positivity and compact support, to provide a robust calibration and
imaging method which is strongly convergent in practice \citep{tho01}.

An analytic solution for the fidelity of the reconstructed image is
not tractable given the complex instrumental model and the common use
of non-linear deconvolution methods. A method for quantitative
fidelity assessment is, however, of vital scientific importance in
radio interferometry. The increasing instrumental complexity and
output data rates of current and future radio interferometers, such as
the Atacama Large Millimeter Array
(ALMA\footnote{http://www.alma.nrao.edu}), the pre-cursor Combined
Array for Research in Millimeter-wave Astronomy
(CARMA\footnote{http://www.mmarray.org}), the Expanded VLA Project
(EVLA\footnote{http://www.aoc.nrao.edu/evla}), and the Square
Kilometer Array (SKA\footnote{http://www.skatelescope.org}), require
automated pipelined data reduction, as opposed to custom interactive
reduction, if these telescopes are to achieve their full scientific
potential and be accessible to the broadest astronomical user
community. This is particularly important for potential observers from
other wavebands, or those who do not have a significant investment in
specialized radio interferometric expertise at their local
institutions. Pipelined data reduction can remove these barriers to
entry, but requires a method for quantitative fidelity assessment to
allow the statistical significance of features in the final science
images to be clearly delineated. This is also true for general
scientific analysis and interpretation of archived radio data, in
repositories such as the National Virtual Observatory
(NVO\footnote{http://www.us-vo.org}). Apart from their importance in 
pipeline and archive analysis, we also note that fidelity assessment methods
are useful in standard interactive data reduction, particularly
for long-duration targeted imaging observations where high-fidelity is
a specific goal.

Recent advances in techniques in computational statistics and the
availability of leading-edge community resources for high-performance
computing (HPC), allow fundamentally new approaches to be taken to the
problem of radio-interferometric imaging fidelity assessment. We
report here on the investigation of one such modern statistical
technique, namely bootstrap resampling of dependent data
\citep{lah03}, applied to the test case of polarimetric calibration
and imaging of Very Long Baseline Interferometry (VLBI) data. This
test problem has been chosen for study as it is broadly representative
of the calibration and imaging problems in radio interferometry in
general, and is also of intrinsic scientific and technical interest.
It is not, however, the only imaging application in radio
interferometry for which fidelity assessment is scientifically
important. Other such problems include, for example, high-fidelity
imaging of strong, extended sources in total intensity, and we reserve
their study for future work. We note the use of the bootstrap to
assess the functional precision of medical tomography images in a
related inverse imaging problem domain \citep{mai97}.

The advent of the Very Long Baseline Array (VLBA\footnote{The National
Radio Astronomy Observatory is operated by Associated Universities,
Inc., under cooperative agreement with the National Science
Foundation}), which was engineered with a uniform instrumental
polarization response and a higher overall sensitivity than previous
US VLBI arrays \citep{kel85}, has significantly broadened the scope
and range of science possible using VLBI polarimetry\citep{war94}.
Polarization VLBI studies are frequently concerned with imaging weak
polarized emission, as common target sources such as continuum
extra-galactic radio source cores typically have a degree of
integrated linear polarization not exceeding several percent of the
total Stokes $I$ brightness \citep{caw93}. Image fidelity assessment
methods are therefore particularly relevant to VLBI polarimetry
\citep{rob94}. We note that although other emission processes studied
with VLBI, such as maser emission, may have a higher degree of either
linear or circular polarization than continuum synchrotron sources
\citep{kem02}, it remains equally important to be able to assess the
significance of weak polarized emission in imaging studies of these
objects as well.

In this paper we explore the use of the statistical technique of
bootstrap resampling to estimate the pixel-based variance in the
output images produced by the calibration and imaging of simulated
polarization data from the VLBA. These variance images are
inter-compared with results obtained by direct Monte Carlo simulation,
and confirm the validity of the bootstrap approach in this domain.

The paper is structured as follows: Section 2 describes the theory of
polarization calibration and bootstrap resampling. The details of the
methods used in this numerical study are presented in Section 3 and
the numerical results in Section 4. The discussion and conclusions are
presented in Section 5 and Section 6 respectively.

\section{Theory}

\subsection{The imaging equation for VLBI polarimetry}

Radio-interferometric polarimetry, which allows imaging of the radio
brightness in all four Stokes parameters $\{I, Q, U, V\}$, is possible
if a sufficient subset of antennas in the array is equipped with
orthogonal polarization receptors (usually crossed-linear or
oppositely-polarized circular) and the correlator forms all available
polarization cross-products between the independent polarization
channels at each antenna on each baseline. The calibration algebra for
interferometric polarimetry was first derived by \citet{mor64}, with
further work provided by \citet{con69} and \citet{wei73}. The analysis
for the specialization of continuum VLBI polarimetry has been
developed systematically over the past twenty years \citep{cot84,
rob84, rob91,cot93,rob94,lep95,kem96}. A corresponding signal path
analysis for spectral line polarization VLBI, including a formulation
of the calibration and propagation effects along the signal path from
the top of the atmosphere through the antennas, feeds, receiving
electronics and correlator, was provided by \citet{kem93} and
\citet{kem95}.

A general mathematical framework for radio-interferometric polarimetry
is described in a series of papers by \citet{ham96a}, \citet{sau96},
\citet{ham96b} and \citet{ham00}. This series of papers (hereinafter
HBS) parametrizes the signal path analysis at each antenna as the
product of a series of $(2 \times 2)$ Jones matrices, by analogy with
their use in transmission optics. This mathematical formulation
represents no new physics over earlier analyzes of the signal path and
the resultant polarimetric interferometer response. However the use of
Jones matrices and the outer matrix product allows a concise
formulation in which the calibration algebra can be expressed
independent of the choice of a specific polarization receptor basis,
be that linear or circular.

The outer matrix product for two Jones matrices $A$ and $B$, denoted
$A \otimes B$, is cited by HBS as:

\begin{equation}
A \otimes B  = 
 \left(
\begin{array}{cccc}
a_{00}b_{00} & a_{00}b_{01} & a_{01}b_{00} & a_{01}b_{01} \\
a_{00}b_{10} & a_{00}b_{11} & a_{01}b_{10} & a_{01}b_{11} \\
a_{10}b_{00} & a_{10}b_{01} & a_{11}b_{00} & a_{11}b_{01} \\
a_{10}b_{10} & a_{10}b_{11} & a_{11}b_{10} & a_{11}b_{11}
\end{array}     \right) 
\end{equation}

Each quadrant of $A \otimes B$ is formed as the product of the
corresponding element in $A$ times the matrix $B$. The outer product
of two $n \times n$ matrices has dimension $n^2 \times n^2$.

\citep{cor95a} generalized the HBS framework to include image-plane
calibration effects, leading to an imaging equation of the general
form:

\begin{equation}
V_{mn} = \prod_{\kappa} \left[G_m^{\kappa} \otimes G_n^{\kappa*}\right] \int_{\Omega} \prod_{\kappa} \left[T_m^{\kappa}(\vecrho) \otimes T_n^{\kappa*}(\vecrho)\right] e^{-2\pi j \vec{b}_{mn} \cdot (\vecrho - \vec{\rho_s})}\ K\ S(\vecrho) d\Omega \label{eqn-ie}
\end{equation}

where $\otimes$ denotes the outer matrix product, $j=\sqrt{-1}$,
$S(\vecrho)$ is the radio brightness expressed as a Stokes four-vector
toward the unit direction $\vecrho \in \Omega$, and $\Omega$ is the
region of astronomical interest on the sky centered on $\vec{\rho_s}$,

\begin{equation}
S(\vecrho) = \left( \begin{array}{c} 
I \\
Q \\
U \\
V 
\end{array} 
\right) (\vecrho)
\end{equation}

$V_{mn}$ is the measured visibility correlation between antennas $m$
and $n$ in a polarization receptor basis $(p,q) \in \{(X,Y), (R,L)\}$,
for linearly- and circularly-polarized feeds respectively,

\begin{equation}
V_{mn} = \left( \begin{array}{c} 
V^{pp} \\
V^{pq} \\
V^{qp} \\
V^{qq}
\end{array} 
\right)_{mn}
\end{equation}

$G_m^{\kappa}$ is an element in the sub-series product of Jones
matrices which have no direction dependence, $G_m^{\kappa} \neq
f(\vecrho)$, and thus describes visibility-plane signal path
corrections at antenna $m$ of type $\kappa$, where $\kappa$ enumerates
the sub-types of the calibration corrections in the signal
path. Examples of visibility-plane calibration correction sub-types
enumerated by $\kappa$ include those arising from bandpass and
electronic gain corrections. In the image plane, $T_m^{\kappa}$ is an
element in the product of Jones matrices with direction dependence,
$T_m^{\kappa} = f(\vecrho)$, thus describing image-plane corrections
at antenna $m$ of type $\kappa$, and $K$ is a fixed $(4 \times 4)$
conversion matrix which maps the Stokes four-vector $S(\vecrho)$ into
polarization correlation pairs $V^{pq}$ in the receptor polarization
basis.

We omit decorrelation losses (HBS) and baseline-dependent digital
signal processing corrections \citep{cor95a} in this expression, as
they are not relevant to the work presented in this paper. The
individual Jones matrix elements for each correction matrix are
expressed in the adopted polarization receptor basis $(\vec{e_p},
\vec{e_q})$ and can also be parametrized arbitrarily on parameters
$\xi_k$ defining an individual instrumental model:

\begin{equation}
G  = 
 \left(
\begin{array}{cc}
g_{00}(\xi_k) & g_{01}(\xi_k) \\
g_{10}(\xi_k) & g_{11}(\xi_k)
\end{array}     \right) 
\left(
\begin{array}{c}
\vec{e_p} \\
\vec{e_q}
\end{array}
\right)
\end{equation}

The sets of Jones matrices in the visibility-plane, $\{G_m\}$, and the
image-plane, $\{T_m\}$, model the calibration corrections for the full
signal path at each antenna. These individual calibration components
are enumerated in further detail by \citet{cor95a}, \citet{noo95}, and
\citet{noo96}. For the chosen case of VLBI polarimetry used here to
evaluate the bootstrap resampling for radio-interferometric imaging
fidelity assessment, we restrict the instrumental corrections to the
Jones matrices for the parallactic angle correction, $P_m=G_m^P$, and
the instrumental polarization response, $D_m=G_m^D$. These formally
both depend on direction, $\vecrho$, but for the narrow field of view
typical in VLBI we consider these as direction-independent corrections
at the center of the field, with $P_m, D_m \in \{G_m\}$. These are the
on-axis values accordingly. For the VLBA antennas, which have
altitude-azimuth (alt-az) mounts and circularly polarized receptors,

\begin{equation}
P_m  = 
 \left(
\begin{array}{cc}
e^{-j \amt} & 0 \\
0 & e^{j \amt}
\end{array}     \right) 
\end{equation}

where $\amt$ is the time-variable parallactic angle at antenna $m$,
which is known analytically (TMS). The instrumental polarization
response, also known as the $D$-terms or polarization leakage
\citep{mor64,con69}, take the Jones matrix form:

\begin{equation}
D_m  = 
 \left(
\begin{array}{cc}
1 & d_m^p \\
d_m^q & 1 
\end{array}     \right) \label{eqn-dterm}
\end{equation}

where $d_m^{(p,q)}$ are the traditional complex components of the
$D$-term at antenna $m$ for polarization $(p,q)$ \citep{noo95}.

\subsection{Strategies for VLBI instrumental polarization calibration}

By design of the presented work, the VLBI polarimetry data considered
here require no calibration for electronic gain amplitude or
phase. Equivalently stated, there is no diagonal Jones matrix in
$\{G_m\}$, outside of $P_m$ or $D_m$, which needs to be applied to
correct the data. This is the usual condition for polarization VLBI
data in current data reduction practice before the final incremental
solution for the polarization leakage terms. The residual problem of
solving for $D_m$ requires knowledge of $S(\vecrho)$ or a joint
solution for $S(\vecrho)$ and $D_m$ simultaneously, as is evident from
equation~\ref{eqn-ie}. In general, this problem is more difficult for
polarization VLBI than connected-element interferometry, due to the
higher spatial resolution and the relative absence of calibrator
sources with a known polarization structure at milliarcsecond (mas)
resolution. A number of approximations for the source model
polarization structure have been developed in the past to sharply
reduce the number of free parameters $\varkappa_l$ describing the
source polarization model, and thus allow a single fit for the source
model unknowns $\varkappa_l$ and the $D_m$ simultaneously. For an
alt-az array such as the VLBA, each antenna has a sufficiently
different variation of parallactic angle with time to allow these
parameters to be separated. Expressed equivalently, the basis
functions $e^{j\amt}$ are non-degenerate over a sufficient range in
parallactic angle variation.  The source approximation methods in
common VLBI use are summarized by \citet{kem99}. These approximations
include assumptions of: i) a linearly-unpolarized calibrator,
$(Q,U)=0$ \citep{rob84}; ii) an unresolved calibrator,
$(Q,U)=\varkappa$ \citep{cot84}; iii) a similarity approximation,
$Q+jU=\varkappa I$ \citep{cot93}; iv) a multi-component similarity
approximation, $\sum_l (Q_l+jU_l)=\varkappa_l I_l$ \citep{lep95}, and,
v) a spectral-line approximation, $Q(\nu)+jU(\nu)=\varkappa_{\nu}I(\nu)$
\citep{kem97}. This family of solution methods will be referred to as
source approximation methods in what follows. These approximations
reduce the source polarization model parameters to fewer than
required for a fully general polarization model by fixing the scaling
factor $\varkappa$ per image (iii), per Gaussian image component (iv),
or per spectral channel (v) respectively.

It has also been common practice in polarization VLBI to linearize the
feed leakage calibration equations by ignoring terms of order $O(D^2)$
or $O(D \cdot (Q+jU))$ \citep{rob84} and to use only the
resulting linear equations for the cross-polarized visibilities
$(V^{pq}, V^{qp})$ in a linear fit for $\varkappa_i$ and $D_m$
\citep{cot93, kem95, lep95, kem97}. This linearization has also typically
been applied to the equation inverse when correcting the visibility
data for the polarization leakage $D-$terms. The use of the
matrix-based HBS framework implicitly includes all terms without
linearization.

\subsection{Polarization self-calibration}

Polarization self-calibration has been proposed frequently in the
literature as a likely optimal technique for solving simultaneously
for the source model and the instrumental polarization response
\citep{rob94,sau96}. This is by direct analogy with the cornerstone
role self-calibration plays in radio interferometry in allowing a
simultaneous solution for the total intensity distribution and complex
antenna-based electronic gains. A recent review of electronic gain
self-calibration is provided by \citet{cor99}. Several of the source
approximation methods described in the previous section have already
been applied iteratively, as a variant of polarization
self-calibration \citep{kem96, kem97}.

A simple definition of polarization self-calibration is that of an
joint (often iterative) solution for $S(\vecrho)$ and $D_m$ within a
calibration framework for radio-interferometric polarimetry, with
optional constraints applied in the visibility and image planes to
regularize the solution as required. As such, it is the direct analog
of standard total intensity phase and amplitude self-calibration. In
the HBS framework, general self-calibration is simply any joint
solution for a set of unsolved Jones calibration matrices in $\{G_m\}$
and $\{T_m\}$ simultaneously with the source brightness distribution
$S(\vecrho)$; all are free parameters in the imaging equation on a
equal footing. In this nomenclature, polarization self-calibration, as
defined above, is a simple subset of general self-calibration in the
imaging equation formalism.

In the imaging equation framework [\ref{eqn-ie}], given an estimate
for $D_m$ and the known analytic $P_m$, the source brightness can be
determined from ${{\partial \chi^2}\over{\partial S(\vecrho)}}$ and
the use of standard deconvolution techniques, as described by
\citep{cor95b}, where $\chi^2$ is computed as the difference between
the observed and predicted visibility data. Conversely, given an
estimate of $S(\vecrho)$, and the known analytic $P_m$, $\chi^2$ can
be formed at the position of the unknown $D_m$ in the imaging equation
by computing contributions from the right-hand and left-hand sides of
[\ref{eqn-ie}] respectively through to the position of $D_m$ in $\prod [G_m
\otimes G_m^*]$ and forming $\chi^2$ by inserting the current value of
the $D_m$ matrix at that position \citep{cor96}. This also allows a
direct computation of ${{\partial \chi^2}\over{\partial D_m}}$, which
can be used in a non-linear fit for the off-diagonal Jones matrix
elements of $D_m$ in each polarization solution interval. We
re-express this mathematically by defining two operators,
$\mathcal{R}^{\beta}$ and $\mathcal{L}^{\beta}$, which operate from
the right- and left-hand sides of the imaging equation respectively,
through to the position of a visibility calibration component
$G_m^{\beta}$ being solved for. The operators yield 4-vectors
containing the accumulated product of visibility cross-correlations at
that index point in the imaging equation. For baseline $mn$,

\begin{eqnarray}
\mathcal{L}^{\beta}_{mn}  & = & V_{mn} \prod_{\kappa}^{\beta-1} \left [\gma \otimes\gnac \right ]^{-1} \\
\mathcal{R}^{\beta}_{mn} & = & \prod_{\kappa=\beta+1} \left [\gma \otimes \gnac \right ]  \int_{\Omega} \prod_{\kappa} \left[T_m^{\kappa}(\vecrho) \otimes T_n^{\kappa*}(\vecrho)\right] e^{-2\pi j \vec{b}_{mn} \cdot (\vecrho - \vec{\rho_s})}\ K\ S(\vecrho) d\Omega 
\end{eqnarray}

The calibration solver minimizes,

\begin{equation}
\chi^2 = \sum_{mn} \| (G_m^{\beta} \otimes G_n^{*\beta}) \mathcal{R}^{\beta}_{mn} - \mathcal{L}^{\beta}_{mn} \|
\end{equation}

Polarization self-calibration can be achieved in this framework using
an iterative solver which starts with an initial value, $D_m^0$, for
each D Jones matrix, and iterates with optional visibility- or
image-plane constraints to solve successively for $S(\vecrho)$ and the
set of $D_m$ per antenna and polarization solution interval, within
the imaging equation model. The data are pre-averaged in the fit over
time intervals short with respect to the maxima $|{{d
\amt}\over{dt}}|$ and $|{{d S(\vecrho)}\over{dt}}|$ where $t$ is used
here as a parameter along the uv-tracks in the visibility plane. We
note that $\chi^2$ is formed here from the visibility 4-vectors,
including cross-polarized correlations. The imaging deconvolution
similarly uses the 4-vector of full polarization intensities.

\subsection{Imaging fidelity assessment}

The visibility data measured at loci in the $uv$-plane for each
projected baseline $mn$ constitute a time-series of complex observed
data, denoted here as: $\vobs$. We adopt a simple additive noise
model for the measured visibility data:

\begin{equation}
V_{mn}^{obs} = V_{mn} + \mathcal{N}
\end{equation}

where $V_{mn}$ is given  by equation [\ref{eqn-ie}], and $\mathcal{N}$
is a random  phasor drawn from the complex  normal probability density
for   $\mathcal{CN}(0,\sigma_{th}^2)$  of   zero  mean   and  variance
$\sigma_{th}^2$.  We  adopt   the  statistical  nomenclature  here  of
\citet{kay93} for complex random  variables. The noise contribution is
independent   and   identically  distributed   (IID).   For  a   radio
interferometer   comprised  of  uniform   array  elements   of  system
equivalent  flux  density SEFD  (a  fundamental  parameter of  antenna
sensitivity), bandwidth $\triangle \nu$, and a sample integration time
$\triangle t$, the thermal noise variance is given in the form (TMS):

\begin{equation}
\sigma_{th}^2 = \frac{SEFD^2}{2 \triangle \nu \triangle t}
\end{equation}

The complete set of measured visibility data $\vobs$ can be regarded
as a single statistical realization of a sequence of random variables
$\{\mv_1, \mv_2,...,\mv_N\}$ of sample size $N$ from a joint
multivariate probability density function (PDF),

\begin{equation}
\pv = \frac{1}{\pi^N \sigma_{th}^{2N}}
e^{-[(\mv - V_{mn})^H C_{\mv}^{-1} (\mv - V_{mn})]} \label{eqn-pdf}
\end{equation}

where the covariance matrix $C_{\mv} = \sigma_{th}^2 U$, superscript
${}^H$ denotes conjugate transpose \citep{kay93}, $U$ is an $(N \times
N)$ unit diagonal matrix, and the imaging equation [\ref{eqn-ie}]
yields $V_{mn} = f(\ip,\gma,\tma)$. Although this is a parametric
distribution, here in terms of $(\ip,\gma,\tma)$, these quantities are
not known at the time of observation, and are determined by the
calibration and imaging process used in data reduction. We consider
any solvers for the calibration matrices $(\gma,\tma)$ or the source
brightness distribution $\ip$ as statistical point estimators for
these parameters and denote these estimators as $\hip,\hgma$, and
$\htma$. This perspective connects the observed visibility data
$\vobs$ and the unknown source brightness distribution and calibration
parameters appearing in the PDF, in the form of a standard statistical
inference problem. In this context, the problem of imaging fidelity
assessment is equivalent to that of determining the sampling
distribution properties of the estimators $\hip, \hgma$, and
$\htma$. The imaging problem is assumed regularized, and therefore
convergent and robust, in this analysis, in order to remove the poor
conditioning implicit in the sparse sampling in the $uv$-plane. We
omit the explicit choice of basis representation for $\ip$; this could
be pixel-based or of functional form.

We denote the sampling distribution of $\hip$ as $\fip$. As
the parent distribution $\pv$ for the observed visibility data is
unknown, the sampling distribution $\fip$ is also unknown a priori.
We are physically and logistically constrained from measuring $\pv$ by
drawing all potential realizations of the visibility time series from
the parent population; each observing run is unique in terms of the
instrumental and atmospheric conditions, quite apart from the
logistical impracticality of this approach. In addition, if the
calibration corrections are imperfectly modeled in any significant
sense, $\gma = f(\xi_l)$, the parent PDF will differ in mathematical
form from the assumed model $\pv$. Errors in the form of the
parametric model compromise the statistical inference based on that
model.

\subsection{Bootstrap resampling}

Bootstrap resampling techniques offer an alternative statistical
inference method which is not as sensitive to errors in the assumed
model and does not require that the problem be analytically
tractable. This method constructs statistical resamples from the
measured data realization which mirror the statistical properties of
the unknown parent population. Developed by \citet{efr79} for IID
data, this method has proved to be a powerful technique in modern
computational statistics, with increasingly broad applicability. It
has undergone significant theoretical development since its inception
and modern reviews are provided by \citet{sha95}, \citet{dav97},
\citet{che99}, and \citet{lah03}.  Bootstrap techniques have also been
extended to the case of statistically dependent data which are not
IID. This question is considered in detail in a monograph by
\citet{lah03}, and is of particular importance for our study. The
observed visibility data in radio interferometry are generated by a
process with long-range statistical dependence, defined to first order
by the Fourier transform of the source brightness distribution to the
visibility $uv$-plane [\ref{eqn-ie}]. Other bootstrap resampling
techniques, such as the jackknife and general subsampling methods
\citep{sha95,pol99}, are also relevant to resampling of dependent data
\citep{lah03}.

A single measured realization $\vobs$ of the random process can be
used to construct a distribution function $\tpv$ from which
statistical resamples $\mv^*$ can be drawn. The bootstrap principle
requires that the statistical relationship between $\mv^*$ and $\tpv$
mirror that between $\mv$ and $\pv$ \citep{lah03}. We consider the
specific choice for $\tpv$ in further detail below.

We denote resamples drawn from $\tpv$ conditional on the observed data
$\vobs$ as $\vobs_{1..N_s}^*$; when acted on by the
imaging estimator $\hip$, they yield $S_N^*$. The bootstrap estimate of the
sampling distribution, $\hfip$ can then be obtained from the
distribution of $S^*_N$.

The primary statistical requirement on successful use of the bootstrap
method is that the relationship between the resamples and the PDF
derived from the data realization accurately
reflect the corresponding relationship between the parent population
and the observed data. For IID data, the empirical probability
distribution function is an optimal choice: $\tilde{p}(\bullet) = N^{-1}
\sum_{k=1}^N \| (\mv \leq \bullet)$, where the indicator function
$\|(A)=1$ if $A$ is true, else zero \citep{lah03}. For IID data,
resamples can be drawn from the empirical distribution function
directly, with replacement, to generate a valid bootstrap ensemble
\citep{dav97}.

The IID bootstrap is not statistically valid, however, for the case of
dependent data \citet{sin81}, and modifications to address dependence
are reviewed by \citep{lah03}. We summarize them here in several broad
categories: (i) block bootstraps; (ii) model-based bootstraps, and;
(iii) subsampling bootstraps. Block bootstrap methods resample the
data in blocks of sufficient length to capture the bulk of the
auto-covariance dependence information in the data. Model-based
bootstrap methods model and remove the statistical dependence from the
data and apply the direct bootstrap to the residuals, assumed then to
be IID \citep{lah03}. In this bootstrap method, the model fit is
tailored to the process generating the data. Subsampling draws
unmodified fractional resamples from the data; as such it is a
generalization of the delete-d jackknife \citep{pol99}. Subsampling
preserves directly the long-range statistical dependence in the resamples.

Although the bootstrap is a computationally intensive technique,
geometric advances in available HPC resources make this technique
readily applicable as an approach to radio interferometry imaging
fidelity assessment. We discuss the practical evaluation of this
technique with radio interferometry data in subsequent sections.

\section{Simulation methods}

We have conducted several numerical studies to inter-compare the
results of bootstrap resampling with those obtained by direct Monte
Carlo simulation, as a means to evaluate the applicability and
usefulness of bootstrap resampling techniques in radio
interferometric imaging fidelity assessment. As noted in the
introduction, we have chosen VLBI polarization self-calibration and
imaging as the test case for evaluation, both as it is broadly
representative of general calibration and imaging in radio
interferometry, and also because of its intrinsic scientific and
technical interest. In the language and formulation of Section 2.4, we
seek to measure the sampling distribution properties of the imaging
estimator using several candidate bootstrap resampling techniques, in
order to assess their statistical performance in this problem domain.

\subsection{Polarization self-calibration heuristic}

For the numerical studies conducted here, we have developed a
polarization self-calibration algorithm based on the general
principles outlined in Section 2.3. We do not consider the detailed
scientific optimization of the polarization self-calibration algorithm
here, as this aspect is not central to our study of the bootstrap
method. We require only that the algorithm be broadly representative
of typical calibration and imaging data reduction processes in radio
interferometry.

We adopt a simple polarization self-calibration heuristic accordingly,
starting with an initial unit diagonal D Jones matrix at each antenna,

\begin{equation}
D_m^0  = 
 \left(
\begin{array}{cc}
1 & 0 \\
0 & 1 
\end{array}     \right) 
\end{equation}

and iterating without any visibility- or image-plane constraints to
solve successively for $\ip$ and each $D_m$ per antenna, varying only
the off-diagonal D Jones terms. The data were pre-averaged to 5
seconds by the solver, and the D Jones solutions were assumed to have
no time-dependence. CLEAN deconvolution was used during imaging, with
a stopping threshold equal to the expected thermal noise limit; this
deconvolution threshold was held constant for all self-calibration
cycles. In this implementation, ten iterations of calibration and
imaging were performed using non-progressive self-calibration, in
which the uncorrected observed data were used in each cycle, as
opposed to the corrected data from the previous cycle.

Iterative polarization self-calibration has not been used extensively
in VLBI polarimetry partly due to the lack of software support in
existing data reduction packages. The basic
AIPS++\footnote{http://aips2.nrao.edu} package implements the imaging
equation [\ref{eqn-ie}] and the HBS framework.  For the applications
developed here, we modified a reference version of the AIPS++ code
base\footnote{v1.8 \# 667 from 2003}, held constant for
reproducibility across the course of this work, and used the 
calibration solver and imaging implementation available in the package
at that time. Additional utilities for bootstrap resampling and
analysis were also developed. These capabilites are intended for
future public release in a local package.

\subsection{Fidelity simulations}

\subsubsection{Simulation configuration}

\begin{figure}
\plotone{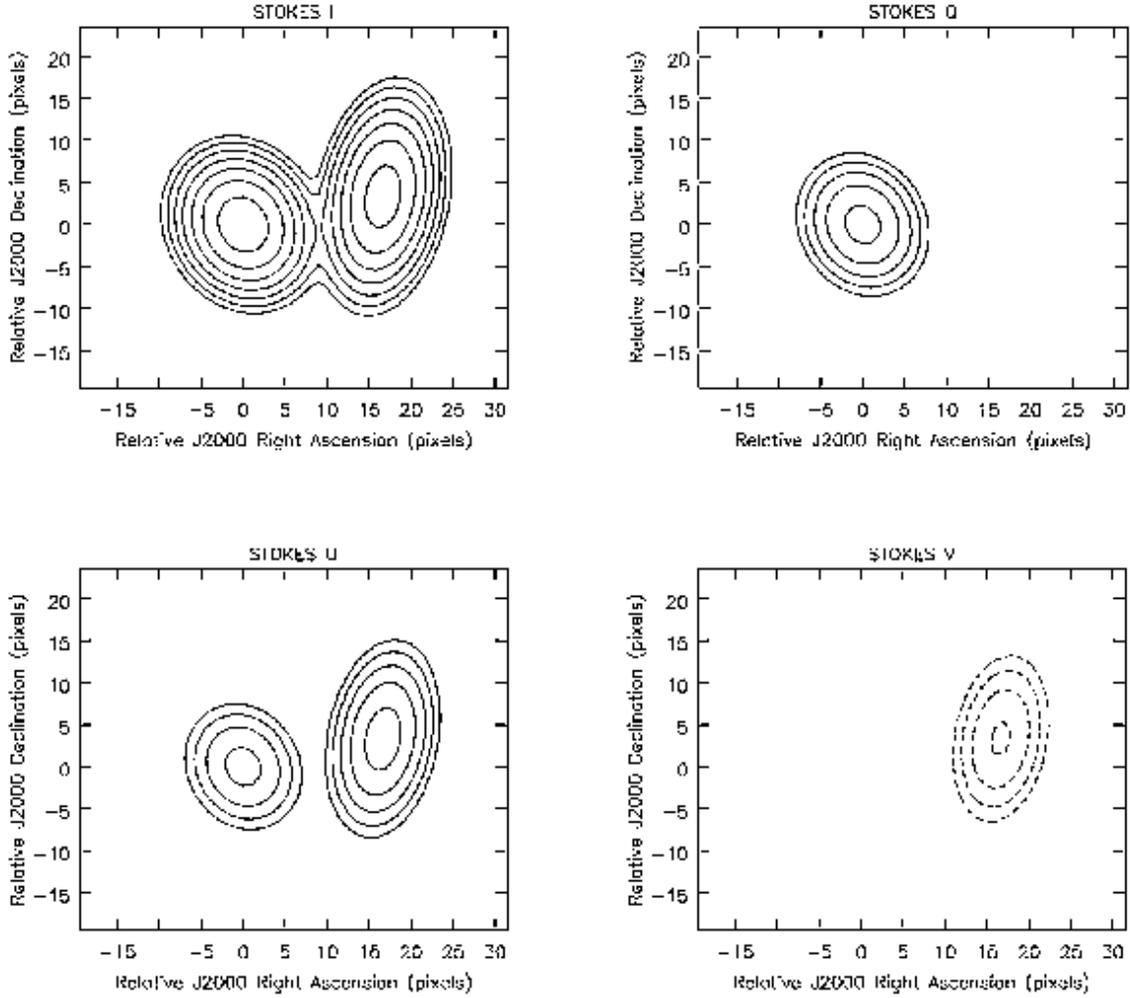}

\caption{Simulation source model, convolved with a circular restoring
beam of 156.007 $\mu as$, plotted in Stokes $\{I,Q,U,V\}$, using
contour levels of $\{-64,\ -32,\ -16,\ -8,\ -4,\ -2,\ -1,\ 1,\ 2,\ 4,\
8,\ 16,\ 32,\ 64\} \times 4.434 \times 10^{-3}$ Jy per beam. The peak
brightness in Stokes $\{I,Q,U,V\}$ is $4.434 \times 10^{-1}$,
$8.867 \times 10^{-2}$, $9.591 \times 10^{-2}$, and $-3.836\times
10^{-2}$ Jy per beam respectively.\label{fig-true-ad}}

\end{figure}

\begin{figure}
\plotone{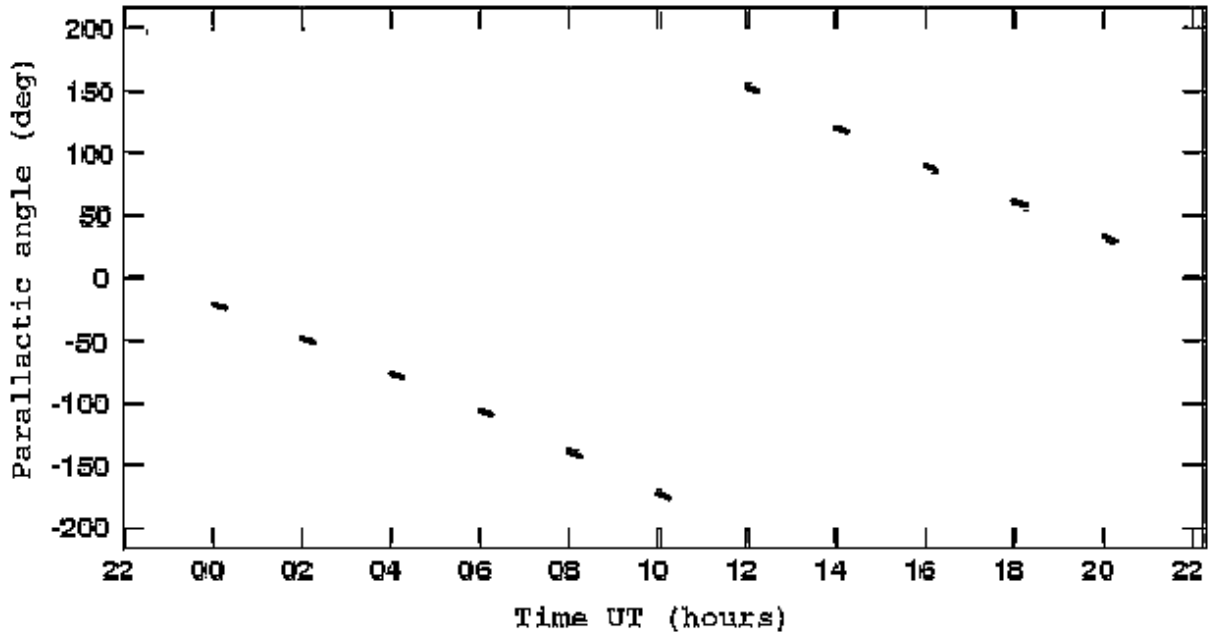}
\caption{Parallactic angle variation at the Pie Town antenna for the data used in the Monte Carlo simulations.\label{fig-parang}}
\end{figure}

For these studies we chose to simulate single-channel data from the
VLBA, at an observing frequency of 43.0 GHz. Data for a source at
$(\alpha_{2000}=12^h,\ \delta_{2000}=+80^{\deg})$ were generated for
all ten VLBA antennas in a set of 15 min scans starting at
$\{t_k=(k-1)^h$ UT$,\ k=1,20\}$, on an adopted date of 1 Feb 2003, using
a correlator integration time $\triangle t=2$ sec. The visibility data
were generated in full polarization in a circular basis $\{RR,LR,RL,LL\}$,
incorporating arbitrarily chosen instrumental D-terms, which are
tabulated for each antenna in table~\ref{tbl-dterms}. The D-terms were
chosen randomly but have a similar mean order of magnitude as those
expected for the VLBA as a whole at this frequency. The source
brightness distribution was chosen randomly as a pair of unit flux
density elliptical Gaussian components matched to the resolution of
the array, with properties summarized in table~\ref{tbl-smodel}, and
plotted in Stokes $\{I,Q,U,V\}$ in figure~\ref{fig-true-ad}. Their
integrated polarization intensities in each polarization were also
chosen at random, but are of the same order as might be observed for
43 GHz SiO maser components using the VLBA. The parallactic angle of the
Pie Town (PT) antenna at the center of the array varied from
-21$^{\circ}$, through -176$^{\circ}$ near transit, to +33$^{\circ}$
at the end of the simulation. This is plotted in
figure~\ref{fig-parang}. The array resolution for this synthesis was
$\sim 150$ $\mu as$ in uniform weighting.  Additive thermal noise was
computed for each integration interval [11], using a nominal SEFD of
1436 Jy published for the VLBA 7mm
band\footnote{http://www.aoc.nrao.edu/vlba/obstatus/obssum/obssum.html}. The
data were generated at an image signal-to-noise ratio (SNR) of
approximately 1300:1, as defined by the ratio of the peak source
brightness to the off-source root-mean-square (rms) noise. This
sensitivity is equivalent to that obtained by observing the unit flux
density source components in an 8 MHz bandwidth, or the same source
components increased in flux density by a factor of 16, in a typical
SiO channel bandwidth of $\frac{1}{32}$ MHz. We adopted a normalized
flux density scale for the source model in this work to allow
different numerical results to be inter-compared readily, and a direct
calculation of the dynamic range as the inverse rms.

\subsubsection{Monte Carlo simulation}

\begin{figure}
\plotone{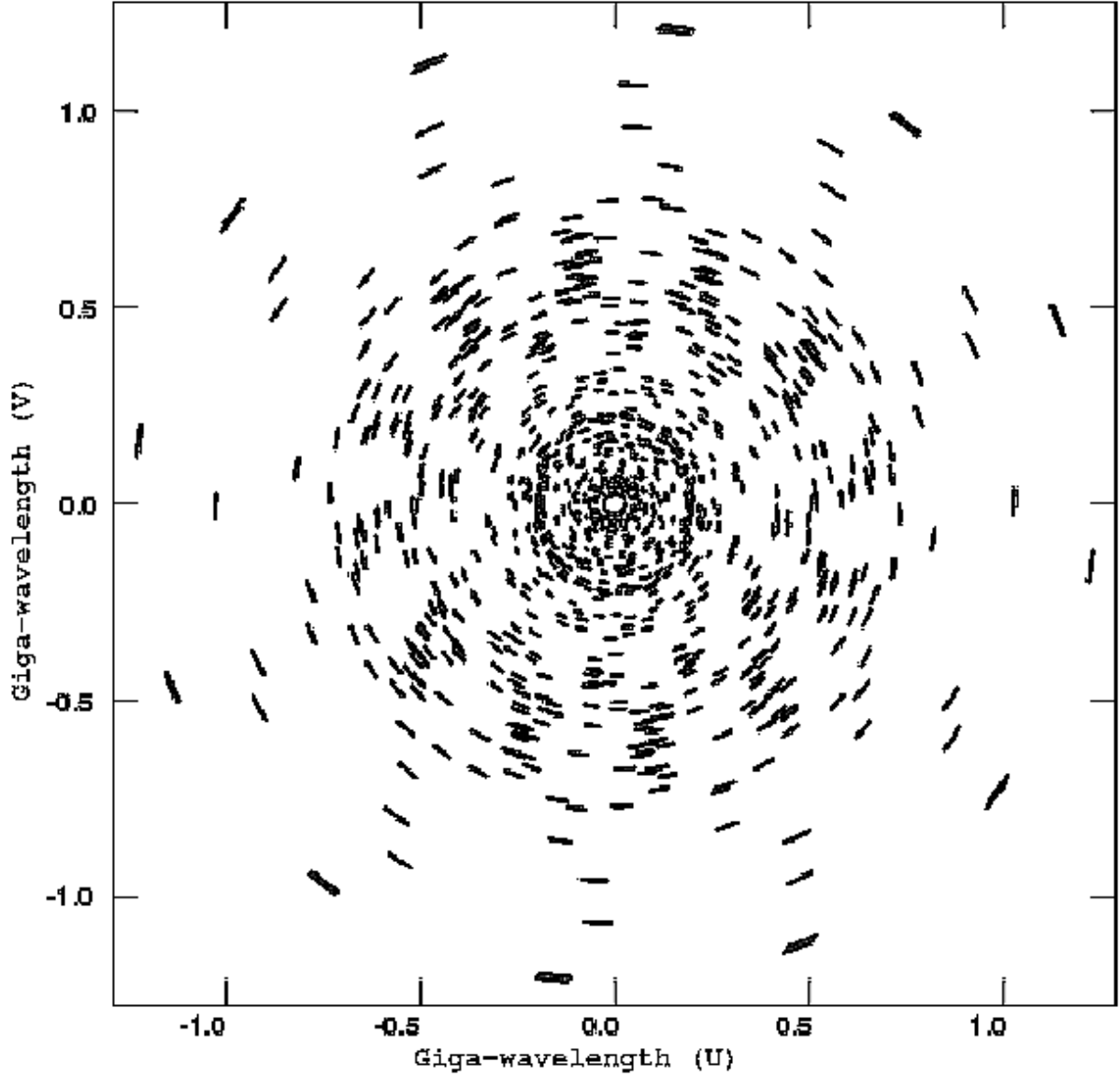}

\caption{$uv$-coverage for the data used in the Monte Carlo
simulations. The simulated source has coordinates
$(\alpha_{2000}=12^h,\ \delta_{2000}=+80^{\deg})$ \label{fig-uvcov}}

\end{figure}

As the first step in the evaluation of bootstrap resampling, a
reference statistical sample containing $N_s=256$ realizations was
generated by Monte Carlo simulation using the parametric model [\ref{eqn-pdf}]
and the array simulation configuration described above. The resulting
Monte Carlo sample consisted of 256 simulated observed visibility time
series data $\vobs^{MC}_{1..N_s}$. The data were generated using a modified
version of the task DTSIM, originally developed by the author in the
AIPS\footnote{http://www.aoc.nrao.edu/aips} package, so that the data
would be generated and reduced in separate packages. Each realization
was reduced using the polarization self-calibration and imaging
heuristic described above, which is equivalent to the application of
the imaging estimator $\hip$ defined in Section 2.4. We chose $\ip$ to
represent the final restored image in a pixel basis $S_{xy}$, and
computed estimator statistics for $\hip$ in this basis over the set of
$N_s$ restored images obtained from application of the imaging estimator to
the full Monte Carlo sample. A fixed circular restoring beam of
full-width half-maximum (FWHM) equal to 156.007 $\mu as$ was used for
all images, set to the geometric mean of the beam in uniform weighting
for the $uv$-coverage of the simulated data, which is shown in
figure~\ref{fig-uvcov}. All images were formed of size $(256 \times
256)$ pixels with a pixel spacing of 30 $\mu as$.

The imaging estimator bias and mean-squared error (MSE) were computed
per pixel with respect to the known source brightness model convolved
with the fixed restoring beam, as shown in
figure~\ref{fig-true-ad}. The mean and variance of the sampling
distribution $\fip$ of the estimator $\hip$, were similarly computed
per pixel over the sample of $N_s$ restored images. The variance was
computed using the approximation $N_s^{-1}\sum (S_{xy}^2) -
{\bar{S}_{xy}}^2$, for computational efficiency. The MSE per pixel was
computed as $N_s^{-1}\sum {(S_{xy} - {\breve{S}_{xy}})}^2$, where
$\breve{S}_{xy}$ is the known true source brightness model convolved
with the fixed restoring beam. Each statistic was computed per Stokes
polarization $\{I,Q,U,V\}$ and per polarization self-calibration
iteration number, across all $N_s$ realizations. Similar estimator
statistics were also accumulated per antenna and self-calibration
iteration number for the polarization calibration estimators
$\hat{G}_m^D$ for each D Jones matrix determined by the polarization
self-calibration and imaging solvers.

The estimator sampling distribution properties estimated using Monte
Carlo simulation were used as the reference in assessing the
statistical performance of the bootstrap techniques. In practical
radio interferometry, direct Monte Carlo simulation of this form is
not practical, as there is no a priori knowledge of the exact source
brightness distribution and calibration matrix values.

\subsubsection{Bootstrap simulations}

We randomly chose the 127$^{th}$ realization $\vobs_{127}^{MC}$ from
the Monte Carlo sample as input to the bootstrap resampling numerical
studies. We resampled from this template realization using both
model-based bootstrap methods and subsampling techniques, and
generated in each case a sample of size $N_s=256$ resampled observed
visibility time series $\vobs^*_{1..N_s}$. The block bootstrap was not
evaluated in this study due to the known long-range dependence in the
data. The polarization self-calibration imaging estimator $\hip$ was
applied to each realization and pixel-based estimator statistics
computed across the $S^*_{xy}$ in the same manner as were computed for
the Monte Carlo sample described above. These statistics estimate the
bootstrap sampling distribution $\hfip$ and can be inter-compared with
those obtained from the Monte Carlo sample to evaluate the statistical
performance of the bootstrap.

For the model-based bootstrap, we adopted a piece-wise polynomial
model for the real and imaginary parts of the observed visibility time
series as a model of the long-term dependence in the data. A model of
the form:

\begin{equation}
\vmn{obs} = \vmod
\end{equation}

was fit over successive bootstrap model intervals $\dtb$ on each
baseline, using least-squares minimization and a conjugate-gradient
solver as implemented in the OptSolve++ package\footnote{OptSolve++ is
distributed by Tech-X corporation (http://www.techxhome.com)}. The
lowest polynomial degree was used in each solution interval which
yielded convergence. The model residuals,

\begin{equation}
\vmn{resid} = \vmn{obs} - \vmod
\end{equation}

were re-centered to avoid introducing bias \citep{lah03, dav97},

\begin{equation}
\vmn{resid'} = \vmn{resid} - N_{\dtb}^{-1} \sum_{\dtb} \vmn{resid}
\end{equation}

and resampled, uniformly and randomly, with replacement, to yield
$\vmn{resid''}$. The bootstrap resample for each $\dtb$ was then
constituted as:

\begin{equation}
\vmn{*} = \vmod + \vmn{resid''}
\end{equation}

Four model-based bootstrap runs were performed, which are labeled M1 through
M4. The bootstrap parameters used for each run are summarized in
table~\ref{tbl-mcode}.

For the subsampling method, visibility points were deleted from the
template realization randomly, avoiding repeated points, until a
specified fraction, $f_s$, of the total number of visibility points in
each sample were removed for each subsample realization
generated. Each subsample realization was processed using the imaging
estimator $\hip$ as used previously, and the same estimator statistics
accumulated over the sample of restored images as were computed for
the model-based bootstrap and Monte Carlo runs. Four subsample
bootstrap runs were performed, labeled S1 through S4. The parameters
used for each subsample run are summarized in table~\ref{tbl-scode}.

\subsubsection{HPC implementation}
All of the preceding Monte Carlo, bootstrap and subsample imaging
estimator runs were computed on the public HPC Linux clusters deployed
as a community resource by NCSA. Many of the bootstrap sample
generation runs were also run in an HPC environment, but some were
executed on a single workstation where possible. The parallel imaging
estimator runs were mapped across the cluster nodes by assigning
subsets of realizations to client nodes, and combining the partial
statistical accumulations from each client node serially on a single
node at the end of the run. A large degree of parallelism was achieved
in these runs due to the limited communication needs between client
nodes in this application. We typically ran over 32-64 cluster nodes,
in individual runs of 4-8 hours duration. The duration depended on the
underlying single-CPU performance of the specified cluster used. We
expect these runs to scale efficiently to larger number of cluster
nodes as they are well-matched to a loosely-coupled HPC architecture
of this type.

\section{Simulation results}

\subsection{Monte Carlo simulations}

\begin{figure}
\plotone{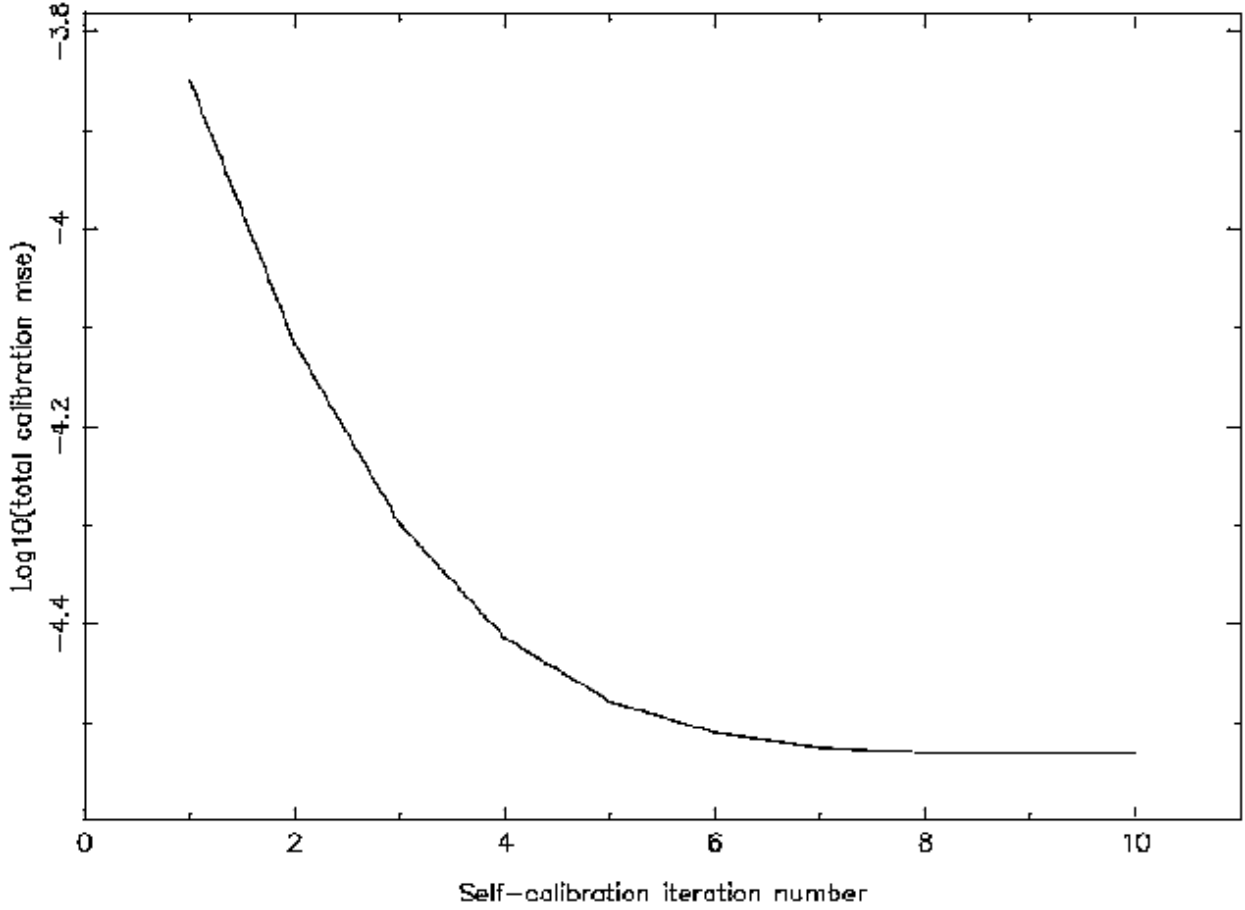}

\caption{Mean-squared error in the off-diagonal D Jones matrix
elements, solved for using polarization self-calibration, averaged
over all antennas in the Monte Carlo simulation, and plotted on a
logarithmic scale against self-calibration iteration number. The total
calibration MSE is computed as: ${{1}\over{2N}} \sum_{p \in (R,L)}
\sum_m^{N} (d_m^p - \breve{d}_{m}^p){(d_m^p - \breve{d}_{m}^p)}^*$, where
$d^p_m$ is the instrumental polarization determined by the solver, and
$\breve{d}_{m}^p$ is the true value used to generate the simulated
data. The instrumental polarization terms are defined in
equation~\ref{eqn-dterm}. \label{fig-calmse}}

\end{figure}

For the Monte Carlo simulations discussed in the previous section, we
plot the total MSE for the calibration estimator $\hat{G}^D_m$, summed
over all antennas $m$, against polarization self-calibration iteration
number in figure~\ref{fig-calmse}.

\begin{figure}
\plotone{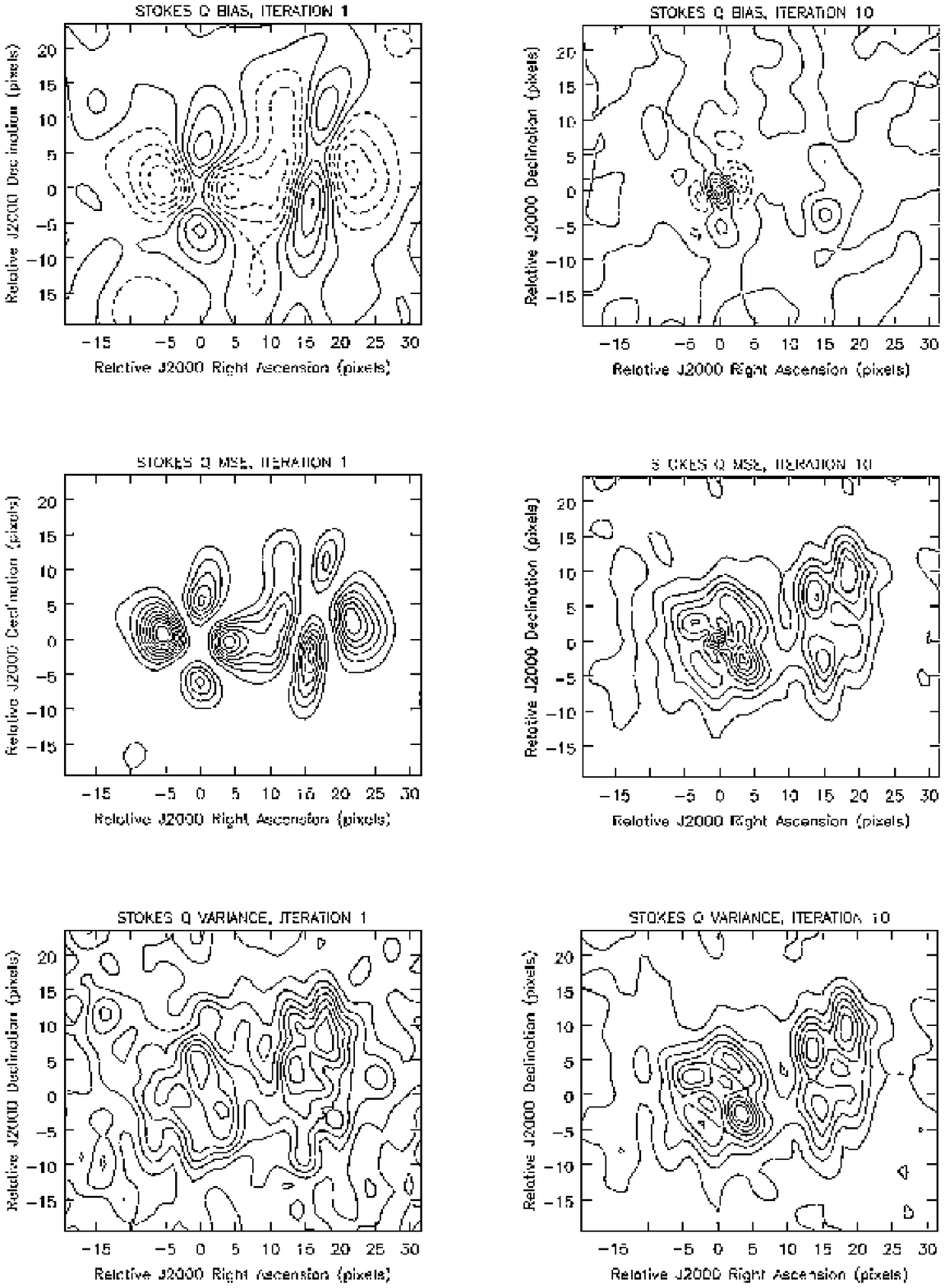}

\caption{The measured imaging estimator bias, mse, and variance for
Stokes $Q$, at polarization self-calibration iteration numbers 1 and
10, for the Monte Carlo sample. Contour levels for the bias plots are
at levels of $\{0.2n,\ n=-5,5\}$ of the local absolute maximum value
($4.566 \times 10^{-3}$ at iteration 1, and $1.009 \times 10^{-3}$ at
iteration 10). Contour levels for the mse and variance plots are at
$\{0.1n,\ n=1,10\}$ of the local maximum. The peak mse is $2.155
\times 10^{-5}$ at iteration 1, and $2.043 \times 10^{-6}$ at
iteration 10. The peak variance is $9.581 \times 10^{-7}$ at iteration
1, and $1.851 \times 10^{-6}$ at iteration 10. All units are in Jy per
beam or (Jy per beam)$^2$\label{fig-mc-q}.}

\end{figure}

\begin{figure}
\plotone{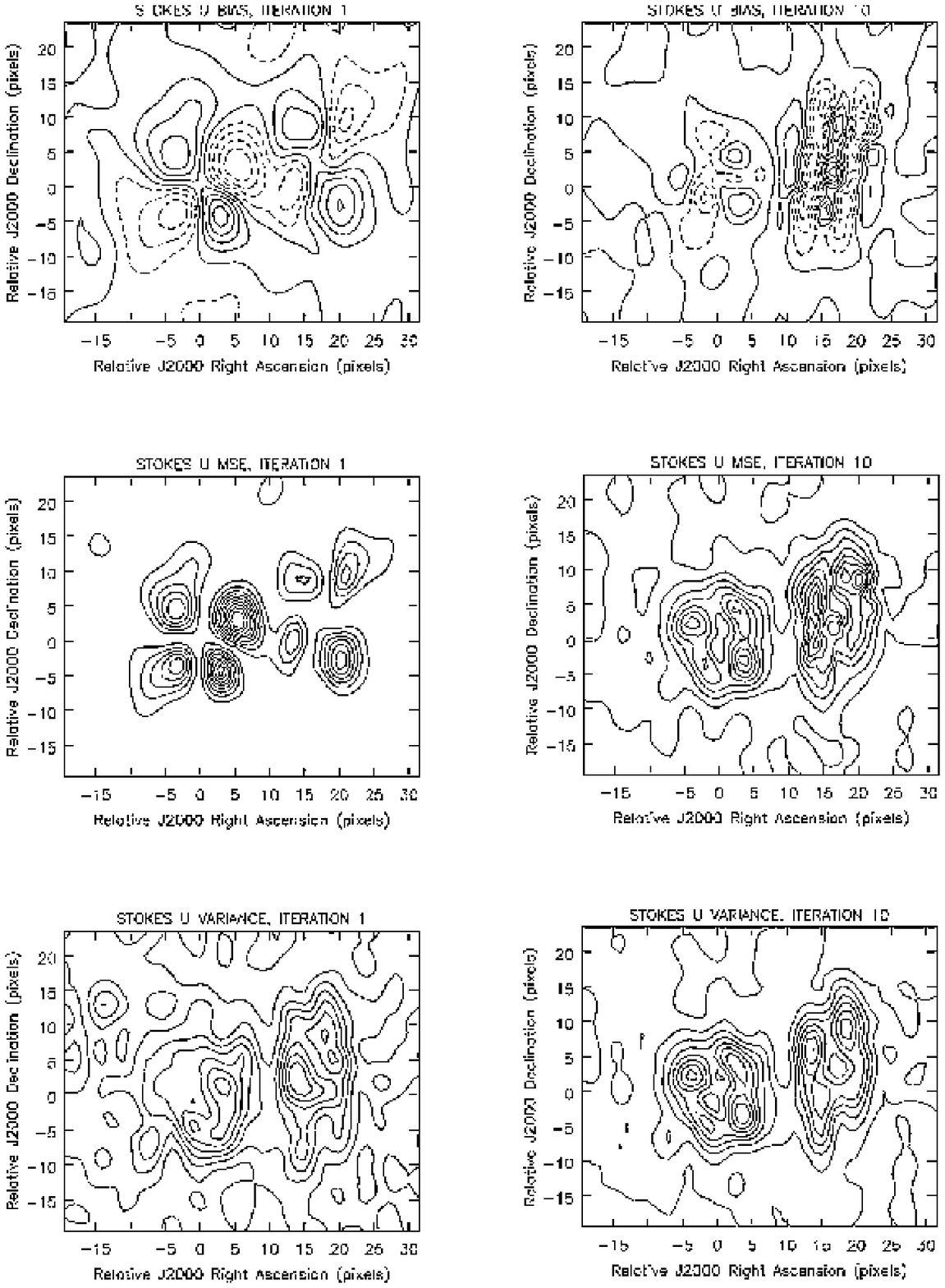}

\caption{The measured imaging estimator bias, mse, and variance for
Stokes $U$, at polarization self-calibration iteration numbers 1 and
10, for the Monte Carlo sample. Contour levels for the bias plots are
at levels of $\{0.2n,\ n=-5,5\}$ of the local absolute maximum value
($4.476 \times 10^{-3}$ at iteration 1, and $8.261 \times 10^{-4}$ at
iteration 10). Contour levels for the mse and variance plots are at
$\{0.1n,\ n=1,10\}$ of the local maximum. The peak mse is $2.082
\times 10^{-5}$ at iteration 1, and $1.730 \times 10^{-6}$ at
iteration 10. The peak variance is $1.041 \times 10^{-6}$ at iteration
1, and $1.543 \times 10^{-6}$ at iteration 10. All units are in Jy per
beam or (Jy per beam)$^2$\label{fig-mc-u}.}

\end{figure}

The MSE, bias, and variance of the pixel-based imaging estimator
$\hip$ are plotted for the restored Stokes $\{Q,U\}$ images, at
self-calibration iteration numbers 1 and 10, in figure~\ref{fig-mc-q}
and figure~\ref{fig-mc-u} respectively.

\subsection{Bootstrap simulations}

To assess the statistical performance of the model-based and
subsampling bootstrap methods, we have computed a goodness-of-fit
statistic which compares the variance image, $var_{xy}$, obtained by
the imaging estimator for a particular bootstrap method against the
corresponding reference variance image derived from Monte Carlo
simulation, $var_{xy}^{MC}$. For the subsample bootstrap, the
estimator statistics need to be scaled by a factor expected to be
proportional to $\sim N(1-f_s)^a$ \citep{dav97}, as derived for the
delete-d jackknife, where the exponent, $a$, is dependent on the
details of the individual model. We estimated the mean scaling factor,
$v_f$, by summing over the inner quarter and over all Stokes
polarizations $\{I,Q,U,V\}$ of the ratio image ,

\begin{equation}
v_f = \frac{1}{N} \sum_{IQUV} \sum_{\Omega_{\frac{1}{4}}} \left ( \frac{var_{xy}}{var_{xy}^{MC}} \right )
\end{equation}

A goodness-of-fit statistic, $v_{MSE}$, was then computed over the same image
region and polarization set, as:

\begin{equation}
v_{MSE} = \frac{1}{N} \sum_{IQUV} \sum_{\Omega_{\frac{1}{4}}} \left ( \frac{var_{xy}}{v_f} - var_{xy}^{MC} \right )^2
\end{equation}

These values are summarized for the model-based and subsample
bootstrap runs in table~\ref{tbl-perform}. For reference, the
bootstrap parameters for the model-based and subsample bootstrap runs
are summarized in table~\ref{tbl-mcode} and table~\ref{tbl-scode}
respectively.

\begin{figure}
\plotone{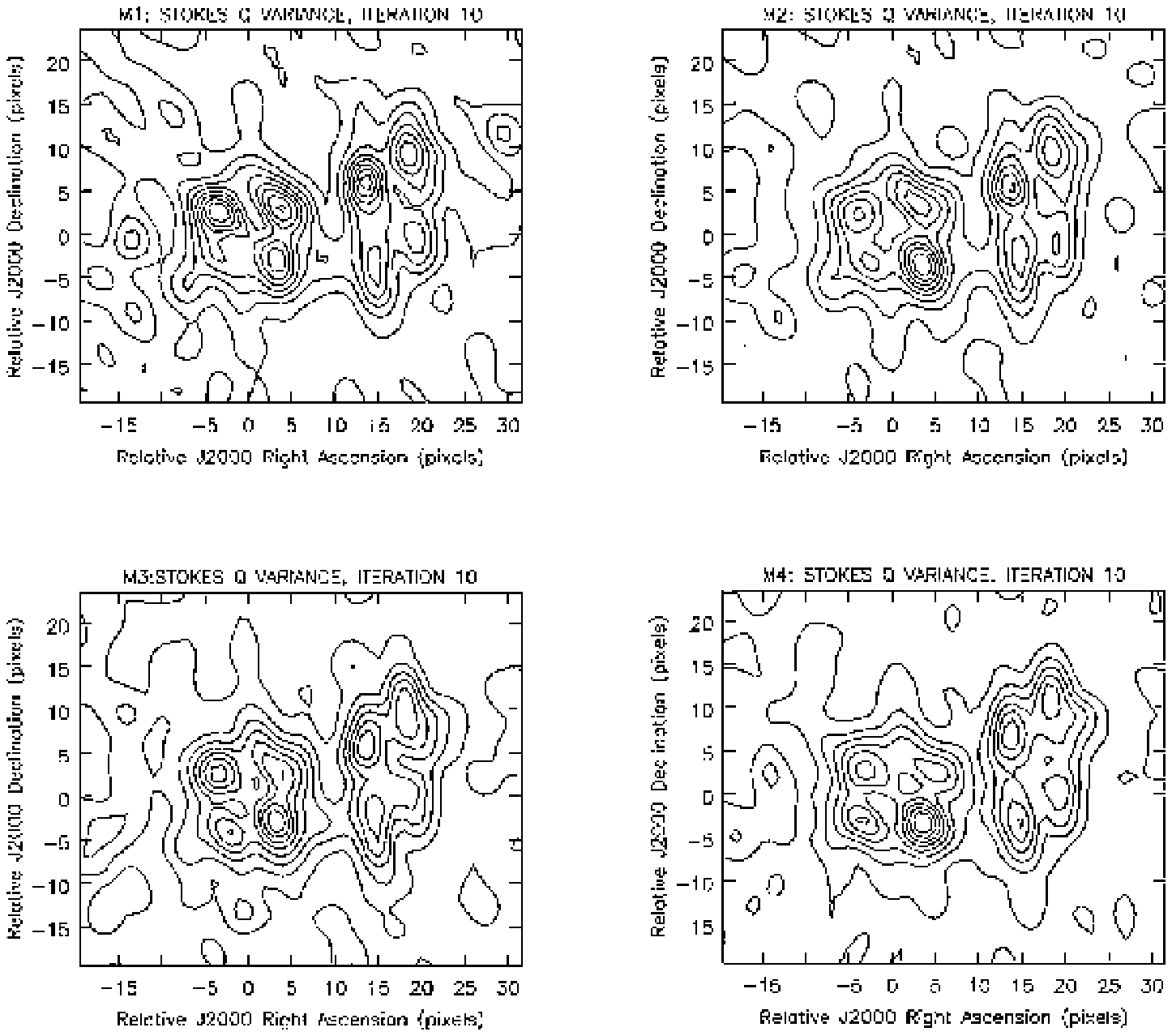}

\caption{The measured imaging estimator Stokes $Q$ variance for the
model-based bootstrap runs M1-M4, at the final polarization
self-calibration iteration number 10. The contour levels are plotted
at $\{0.1n,\ n=1,10\}$ of the local maximum value, which are $1.666
\times 10^{-6}$ (M1), $1.852 \times 10^{-6}$ (M2), $1.758 \times
10^{-6}$ (M3), and $1.855 \times 10^{-6}$ (M4) (Jy per beam)$^2$. The
parameters for the bootstrap run codes M1-M4 are defined in
table~\ref{tbl-mcode}.\label{fig-mb-q}}

\end{figure}

\begin{figure}
\plotone{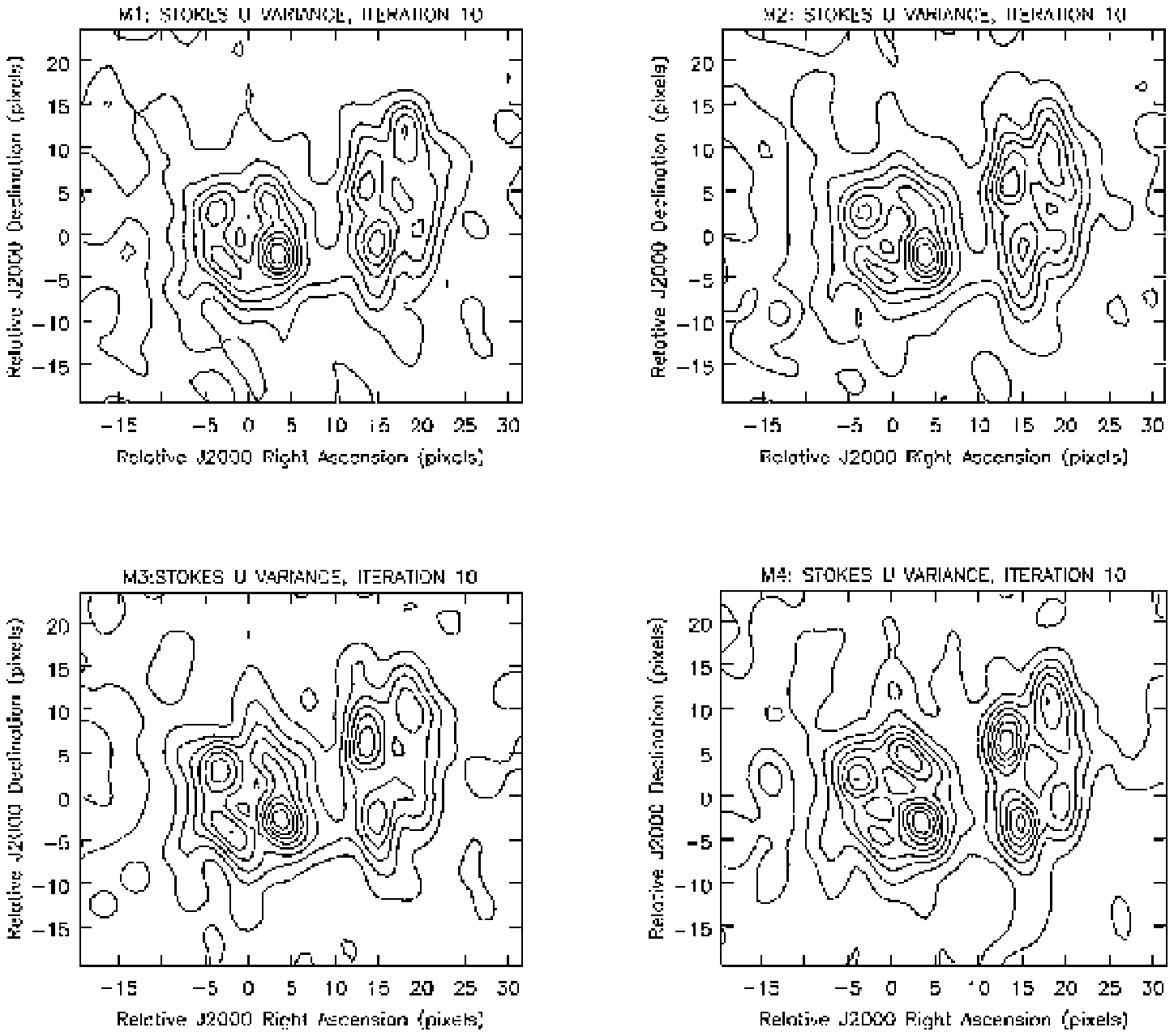}
\caption{The measured imaging estimator Stokes $U$ variance for the model-based
bootstrap runs M1-M4, at the final polarization self-calibration
iteration number 10. The contour levels are plotted at $\{0.1n,\
n=1,10\}$ of the local maximum value, which are $1.737 \times 10^{-6}$
(M1), $1.765 \times 10^{-6}$ (M2), $1.843 \times 10^{-6}$ (M3), and
$1.889 \times 10^{-6}$ (M4) (Jy per beam)$^2$.  The
parameters for the bootstrap run codes M1-M4 are defined in
table~\ref{tbl-mcode}.\label{fig-mb-u}}
\end{figure}

\begin{figure}
\plotone{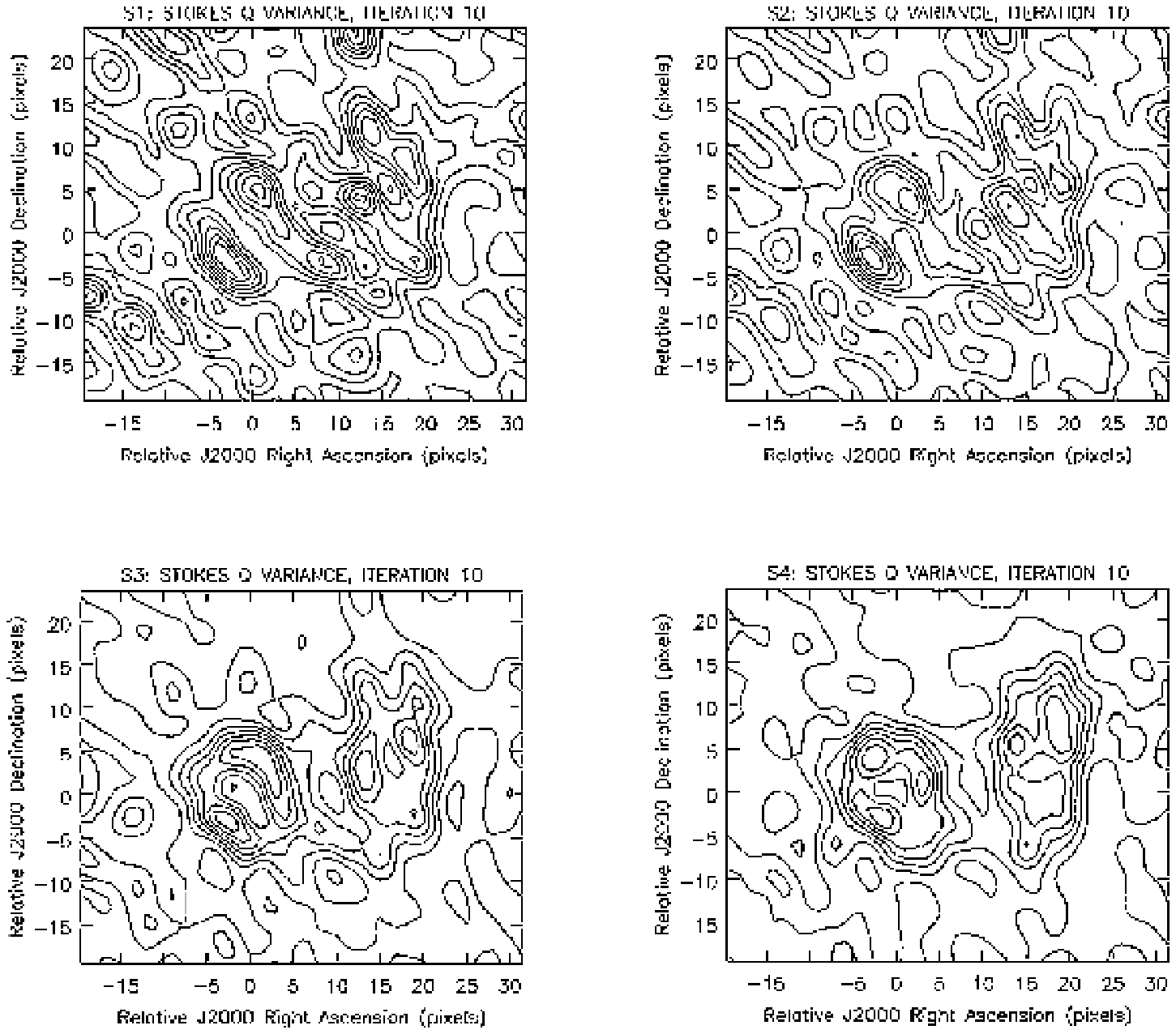}
\caption{The measured imaging estimator Stokes $Q$ variance for the subsample
bootstrap runs S1-S4, at the final polarization self-calibration
iteration number 10. The contour levels are plotted at $\{0.1n,\
n=1,10\}$ of the local maximum value, which are $1.827 \times 10^{-7}$
(S1), $3.849 \times 10^{-7}$ (S2), $7.994 \times 10^{-7}$ (S3), and
$1.890 \times 10^{-6}$ (S4) (Jy per beam)$^2$. The
parameters for the bootstrap run codes S1-S4 are defined in
table~\ref{tbl-scode}.\label{fig-sb-q}}
\end{figure}

\begin{figure}
\plotone{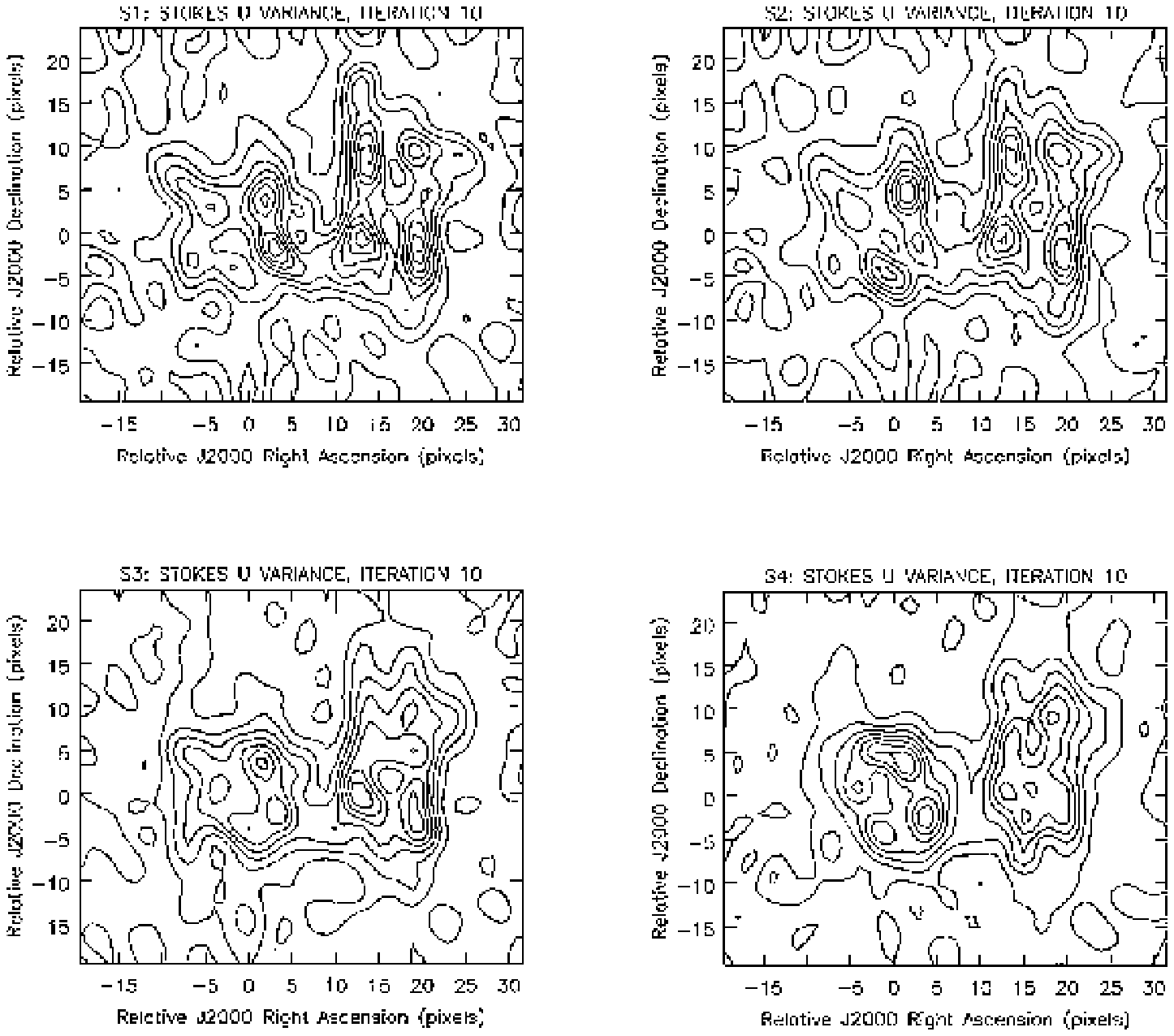}
\caption{The measured imaging estimator Stokes $U$ variance for the subsample
bootstrap runs S1-S4, at the final polarization self-calibration
iteration number 10. The contour levels are plotted at $\{0.1n,\
n=1,10\}$ of the local maximum value, which are $1.817 \times 10^{-7}$
(S1), $3.737 \times 10^{-7}$ (S2), $8.262 \times 10^{-7}$ (S3), and
$2.100 \times 10^{-6}$ (S4) (Jy per beam)$^2$.The
parameters for the bootstrap run codes S1-S4 are defined in
table~\ref{tbl-scode}.\label{fig-sb-u}}
\end{figure}

The variance images in Stokes $\{Q,U\}$ for the final polarization
self-calibration iteration, obtained from the bootstrap runs, are
plotted in figure~\ref{fig-mb-q}, figure~\ref{fig-mb-u},
figure~\ref{fig-sb-q}, and figure~\ref{fig-sb-u}.

\begin{figure}
\plotone{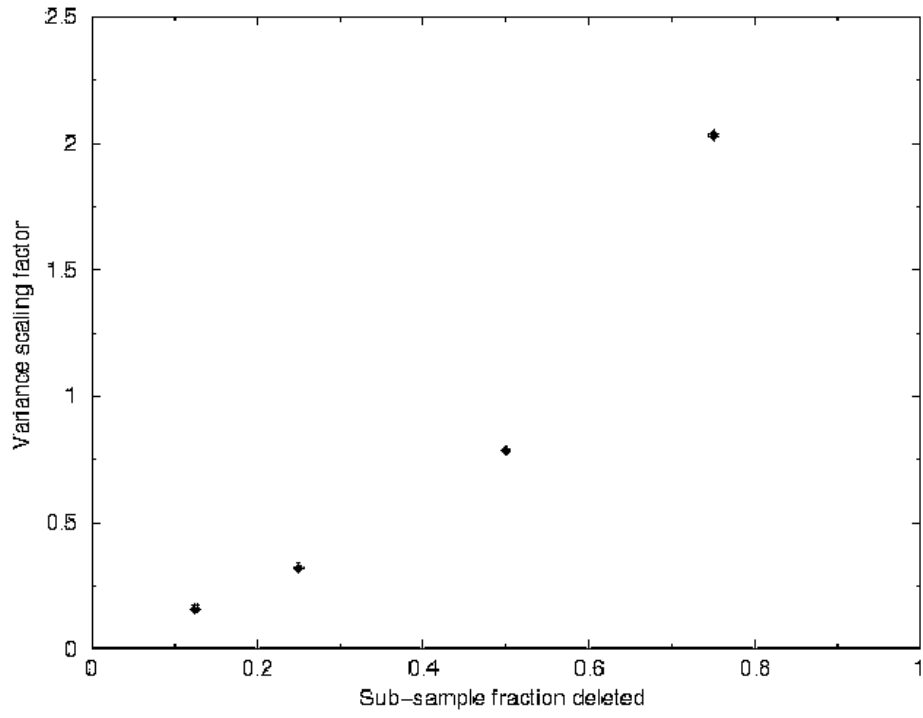}
\caption{The variance scaling factor, $v_f$, plotted against the
subsampling delete fraction, $f_s$\label{fig-vf-fs}.}
\end{figure}

The variation of scaling factor, $v_f$, with subsample delete fraction, $f_s$,
is plotted in figure~\ref{fig-vf-fs}.

\section{Discussion}

The polarization self-calibration algorithm developed as part of this
study of the bootstrap method, has a convergence rate for these data,
shown in figure~\ref{fig-calmse}, that is steeper than exponential as
a function of iteration number. The scope of the current study does
not allow us to extrapolate the quantitative statistical properties of
this calibration estimator to all potential practical applications, but we
believe this to be a viable, general technique for solving for
instrumental polarization in radio interferometry. Most importantly,
it requires no polarization source model approximation, with the
attendant systematic errors which may arise in this case. As is common
for all polarization calibration algorithms in this class, we expect
the solver to be sensitive to the degree of mutual degeneracy of the
parallactic angle basis functions at each antenna
\citep{con69,lep95,sau96}, which can invariably be equivalently
quantified in terms of the maximum range in parallactic angle coverage
for the array as a whole. As noted earlier, the simulated data used in
this study were chosen to have good coverage in parallactic angle.

Intuitively, we expect rapid convergence for polarization
self-calibration algorithms from considering simple information
arguments. For an interferometric observation of $\ntt$
self-calibration intervals, using an array of $\nant$ elements, the
number of self-calibration unknowns is $\ntt \nant$, and the number of
knowns, defined by the data, is $\frac{\nt \nant(\nant-1)}{2}$, where
$N_t$ is the number of unique time-stamps in the observed visibility
data. The ratio of knowns to unknowns is therefore $r = \frac{\nt
(\nant - 1)}{2 \ntt}$. For conventional amplitude and phase
self-calibration, $\nt \sim \ntt$, and $r \sim \frac{\nant}{2}$. For
polarization self-calibration, the D-terms have slowly-varying
time-dependence, $\ntt \sim 1$, and $r \sim \frac{\nt \nant}{2}$. This
ratio is significantly higher than for amplitude and phase
self-calibration and we expect a steeper rate of convergence as a
result. As a corollary, given that the number of free parameters being
solved for in polarization self-calibration is small relative to the
number of visibility data points, it is more difficult to absorb
coherent source brightness distribution errors in the calibration
corrections than in conventional amplitude and phase
self-calibration. This is dependent on the range of parallactic angle
coverage however.

This paper has considered the common case of constant,
visibility-plane polarization calibration, but this polarization
self-calibration algorithm can, in principle, be used to solve for
time- and direction-dependent instrumental polarization corrections.
This is required for high-fidelity, wide-field imaging studies but is
not a routine mode of radio interferometer calibration at present. This
is a subject for future study but we anticipate that the algorithm
performance in this case will depend on the accuracy and number of
free parameters in the parametrization adopted for the time- and
direction-dependence of the D Jones polarization corrections.

Further information on the properties of the polarization
self-calibration algorithm can be obtained by examining the imaging
estimator statistics for the Monte Carlo sample, plotted in
figure~\ref{fig-mc-q} and figure~\ref{fig-mc-u}. At the first
iteration, D Jones matrix errors dominate, and the magnitude of the
estimator bias is higher across the image as a result. At this
iteration, the imaging estimator MSE is dominated by the estimator
bias. At the final self-calibration iteration the D Jones matrix
corrections are better constrained, and the bias is accordingly
reduced. At this point the imaging estimator MSE is dominated by the
estimator variance. Both calibration and deconvolution errors
contribute to the measured imaging estimator statistics. To first
order, errors in the estimated D Jones calibration corrections corrupt
the cross-polarized visibility data with proportional fractions of the
Stokes $I$ visibility \citep{rob94}, so the variance in Stokes $Q$ and
$U$ does not scale linearly with the underlying Stokes $Q$ or $U$
intensity at each pixel.

The statistical performance of the model-based bootstrap, over runs M1
to M4, can be assessed by examining the goodness-of-fit statistic,
$v_{MSE}$, tabulated in table~\ref{tbl-perform}, as well as the
imaging estimator statistics plotted in figure~\ref{fig-mb-q} and
figure~\ref{fig-mb-u}. We expect optimal performance for the
model-based bootstrap when the bootstrap model interval, $\dtb$, is
sufficiently long for adequate SNR in the fit to the bootstrap model
parameters, but not so long as to invalidate the functional model for
the statistical dependence in the data. The goodness-of-fit statistics
listed in table~\ref{tbl-perform}, which capture the degree to which
the bootstrap imaging variance matches that obtained from the
reference Monte Carlo sample, suggest that run M3 reflects the optimal
choice of model-based bootstrap run parameters for the data used in
this study. Run M3 has the lowest $v_{MSE}$; therefore its variance
estimates have the best overall fit to the image variance obtained by
direct Monte Carlo simulation.

\begin{figure}
\plotone{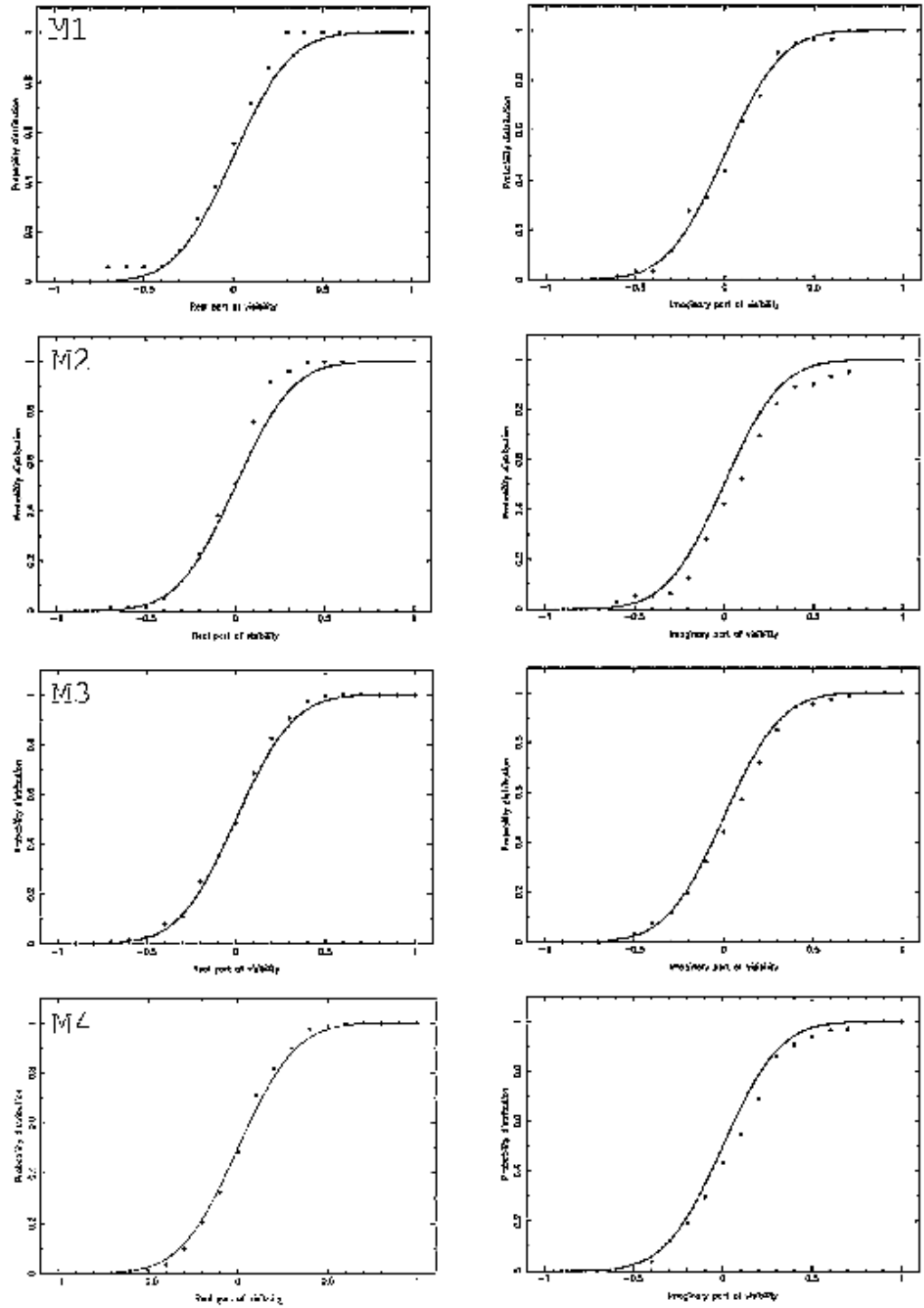}
\caption{The empirical CDF for visibility row 1000, polarization
cross-correlation $LR$, for the model-based bootstrap resamples M1 to
M4. The CDF for the real part of the visibility data is plotted in the
left column, and the CDF for the imaginary part of the visibility data
in right column. The solid line is the expected normal CDF from the
parent distribution. All data have been centered by subtracting the
expected visibility value $V_{mn}$ at this visibility row number
\label{fig-mb-dist}.}
\end{figure}

\begin{figure}
\plotone{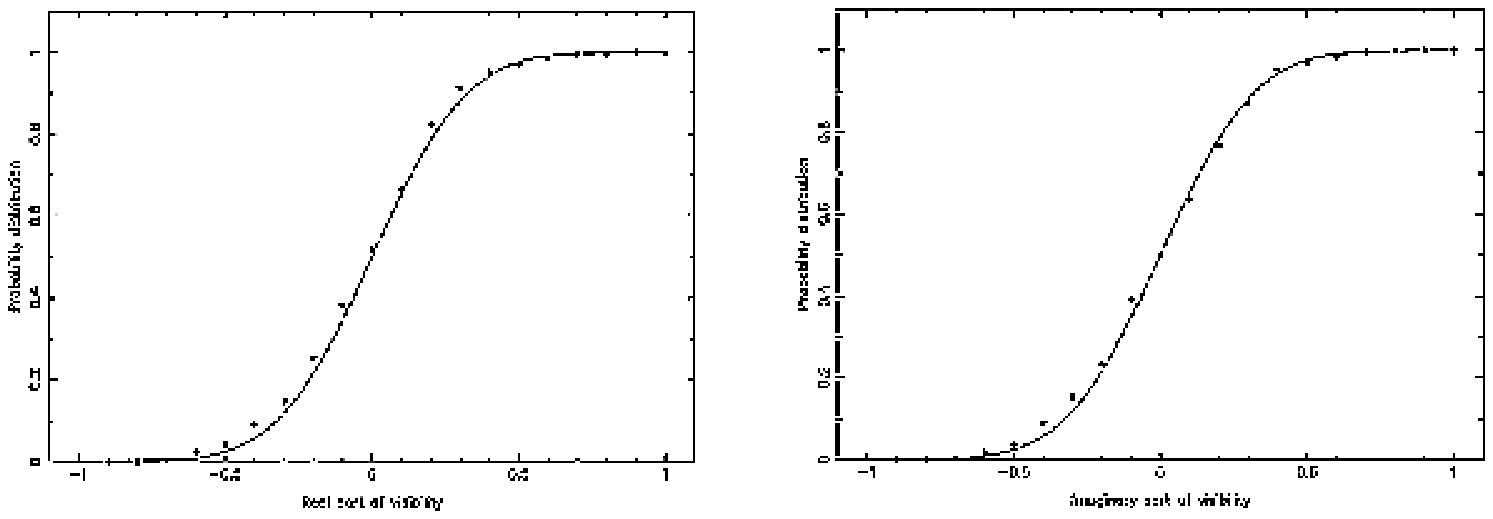}
\caption{The empirical CDF for visibility row 1000, polarization
cross-correlation $LR$, for the Monte Carlo sample. The CDF for the
real part of the visibility data is plotted in the left column, and
the CDF for the imaginary part of the visibility data in the right
column. The solid line is the expected normal CDF from the parent
distribution. All data have been centered by subtracting the expected
visibility value $V_{mn}$ at this visibility row
number\label{fig-mc-dist}.}
\end{figure}

It is clear, however, both from the tabulated goodness-of-fit
statistic $v_{MSE}$, and the imaging estimator statistics, that the
model-based bootstrap performs well over a fairly broad range of
$\dtb$ and $N_p$; the values of these parameters used for the runs M1
to M4 are tabulated in table~\ref{tbl-mcode}. We find this technique
to be relatively robust with respect to the detailed bootstrap model
assumptions, and therefore easy to use in practice. This robustness
over a wide range of tuning parameters is a significant advantage of
the model-based bootstrap. There are, however, differences between the
model-based bootstrap resamples, M1 to M4. As this is a numerical
simulation study, we know the parent population distribution
[\ref{eqn-pdf}], and can compare the empirical distribution of the
resampled data against the known parent distribution. The closer the
match the better the performance of the underlying bootstrap
method. We plot this in the form of a cumulative distribution function
(CDF) comparison in figure~\ref{fig-mb-dist}, for the real and
imaginary parts of the data separately, for a randomly chosen
visibility row number 1000, in polarization cross-correlation
$LR$. The empirical CDF for the reference Monte Carlo sample is
plotted in figure~\ref{fig-mc-dist}. We denote the empirical CDF for
the resampled data as $\tilde{c}^{(R,I)}(\mv)$, and the true parent
CDF as $c^{(R,I)}(\mv)$, after removal of the known value of $V_{mn}$
at this visibility point, and where,

\begin{equation}
c^{(R,I)}(\mv^{(R,I)}) = \frac{1}{2}(1+{\rm erf}(\frac{\mv^{(R,I)}}{\sqrt{2}\ \sigma_{th}}))
\end{equation}

The real and imaginary parts of the visibility data are denoted by
superscript $(R,I)$ respectively.

A goodness-of-fit statistic can be computed in the form $c_{MSE}^{(R,L)} =
N^{-1} \sum^N (\tilde{c}^{(R,I)} - c^{(R,I)})^2$, and we tabulate the
geometric mean of the value for the real and imaginary parts of the
data $c_{MSE} = \sqrt{c^R_{MSE}\ c^I_{MSE}}$ in
table~\ref{tbl-cdf}. This analysis demonstrates that the bootstrap
resamples generated for run M3 match most closely the parent
distribution, consistent with our earlier result concerning the
relative performance of the model-based bootstraps used in the current
study.

We note that the model-based bootstrap does require a detection of the
bootstrap model in the baseline-based bootstrap model intervals
$\dtb$. This is broadly equivalent to the requirement that the data
have sufficient SNR to allow conventional amplitude and phase
self-calibration over a comparable interval. The bootstrap model is
local to each interval $\dtb$ however, and is not equivalent to an
overall model-fit to the source brightness distribution and
calibration corrections for the data as a whole.

The statistical performance of the subsample bootstrap runs, S1 to S4,
can be determined by examination of the goodness-of-fit statistic $v_{MSE}$ is
table~\ref{tbl-perform}, and the imaging estimator statistics plotted
in figure~\ref{fig-sb-q} and figure~\ref{fig-sb-u}. From these results
it is clear that the statistical performance of this bootstrap is more
sensitive to the bootstrap run parameters, here defined by the subsample
delete fraction $f_s$, than the model-based bootstrap. Improved
statistical performance is obtained for $f_s > \frac{1}{2}$ for the
data used in this numerical study. This is consistent with the expected
theoretical properties of the subsample bootstrap, which requires that
$f_s \rightarrow 1$ as the sample size $N \rightarrow \infty$, if the
bootstrap is to remain valid \citep{dav97}.

The variance scaling factor $v_f$ is plotted against subsample delete
fraction $f_s$ in figure~\ref{fig-vf-fs}.  From theoretical
considerations for the delete-d jacknife, we expect $v_f \propto
(1-f_s)^{-a}$ \citep{dav97}. For the data used in this numerical
study, a relation $v_f \propto (1-f_s)^{-2}$ agrees with the data to
first-order, consistent with the expected theoretical result.

Although the subsample bootstrap appears to under-perform the
model-based bootstrap, as assessed by the goodness-of-fit statistic
$v_{MSE}$ in table~\ref{tbl-perform}, this bootstrap has the advantage
of requiring no fit to a local baseline-based bootstrap model and
makes no assumption about the functional form of the long-range
statistical dependence in the data. As such, it is a versatile
technique which can find practical application in
radio-interferometric imaging fidelity assessment.

The computing costs of the both the model-based and subsample
bootstrap fidelity assessment methods are, to first order, given by
the cost of the underlying calibration and imaging algorithm, scaled
by the number of bootstrap resamples, $N_s$. We note above, however,
that this algorithm is highly scalable in parallel computing
environments as each resample can effectively be processed separately.

In this numerical study, we have estimated imaging fidelity by
exploring only the contribution of the thermal noise variation. A
recent Bayesian approach to imaging fidelity assessment for optical
astronomical images is presented by \citet{esc04}.

Bootstrap studies in general require limited statistical model
assumptions and can be used directly with the measured data
realization observed by a radio interferometer array. As such, they
offer significant practical advantages, and we believe them to be a
viable approach to imaging fidelity assessment in the automated pipeline
reduction of data from current and future radio interferometers.

\section{Conclusions}

On the basis of this numerical study, we draw the initial conclusion
that modern resampling techniques in computational statistics offer
significant promise as imaging fidelity assessment techniques for
common calibration and imaging processes used in radio
interferometry. This conclusion applies both to the model-based bootstrap and subsample bootstrap methods. This computationally intensive approach
is now tenable as a result of the recent geometric advances in
available computing resources, and we believe these fidelity
assessment techniques have an important role to play in automated
pipeline reduction environments for modern radio telescopes.

The polarization self-calibration algorithm developed as part of this
study as the framework calibration and imaging test problem, offers a
new approach to interferometric polarization self-calibration which
makes no assumptions regarding the form of the polarization source
model.

Further work is required in assessing the statistical performance of
the polarization self-calibration algorithm and the resampling
techniques over a larger set of simulated data covering a broader 
range of observational parameters.

\acknowledgments
This work was supported by a grant of computing time on the NCSA IA-32
Linux clusters under allocation AST 030025. We thank Drs R. Crutcher
and T. Cornwell for reading the manuscript. We thank the referee for
detailed comments on the manuscript which improved the paper overall.

\clearpage

\begin{deluxetable}{lrrrr}
\tabletypesize{\scriptsize}
\tablecaption{Simulation D-term model values\label{tbl-dterms}}
\tablewidth{0pt}
\tablehead{
\colhead{VLBA antenna} & \colhead{$\|D^R\|$} & \colhead{$\phi^R$ (deg)} &
\colhead{$\|D^L\|$} & \colhead{$\phi^L$ (deg)} \\
}
\startdata
Pie Town, NM      &  0.03 & -57.0 & 0.01 &  23.0 \\
Kitt Peak, AZ     &  0.02 &  40.0 & 0.01 &  -5.0 \\
Los Alamos, NM    &  0.03 &  10.0 & 0.05 &  13.0 \\
Brewster, WA      &  0.04 &  87.0 & 0.02 &   3.0 \\
Fort Davis, TX    &  0.09 & -15.0 & 0.03 &  56.0 \\
St. Croix, VI     &  0.01 & -52.0 & 0.01 &  29.0 \\
North Liberty, IA &  0.05 &  77.0 & 0.01 & -23.0 \\
Owens Valley, CA  &  0.03 &  47.0 & 0.01 &  43.0 \\
Mauna Kea, HI     &  0.03 &  -7.0 & 0.06 &  83.0 \\
Hancock, NM       &  0.03 & -27.0 & 0.02 &  35.0 \\
\enddata
\tablecomments{The D-terms tabulated here in a circular basis translate to a D Jones matrix for antenna $m$ of the form:
\begin{equation}
D_m  = 
 \left(
\begin{array}{cc}
1 & \|D^R\|\ e^{j\phi^R} \\
\|D^L\|\ e^{j\phi^L} & 1 
\end{array}     \right) 
\end{equation}
}
\end{deluxetable}

\begin{deluxetable}{lccccccccc}
\tabletypesize{\scriptsize}
\tablecaption{Simulation source model parameters\label{tbl-smodel}}
\tablewidth{0pt}
\tablehead{
\colhead{} & \colhead{$\triangle \alpha$} 
& \colhead{$\triangle \delta$} & \colhead{$b_{maj}$} 
& \colhead{$b_{min}$} &  \colhead{P.A.} 
& \colhead{$I$} & \colhead{$Q$} & \colhead{$U$} & \colhead{$V$} \\
\colhead{} & \colhead{(mas)} & \colhead{(mas)} & \colhead{(mas)} 
& \colhead{(mas)} & \colhead{(deg)} & \colhead{(Jy)} & \colhead{(Jy)} 
& \colhead{(Jy)} & \colhead{(Jy)} 
}
\startdata
1 &    0 &    0 & 0.2 & 0.15 &  30 & 1 & 0.2 &  0.1 &   0 \\
2 & -0.5 & +0.1 & 0.3 & 0.1  & -10 & 1 & 0   & 0.25 & -0.1 \\
\enddata
\end{deluxetable}

\begin{deluxetable}{lcc}
\tabletypesize{\scriptsize}
\tablecaption{Model-based bootstrap run parameters\label{tbl-mcode}}
\tablewidth{0pt}
\tablehead{
\colhead{Run code} & \colhead{max($N_p$)} & \colhead{$\dtb$ (sec)} 
}
\startdata
M1 & 1 & 60 \\
M2 & 1 & 150 \\
M3 & 1 & 300 \\
M4 & 2 & 900 \\
\enddata
\tablecomments{max($N_p$) is the maximum degree of the model polynomial used in each bootstrap model interval $\dtb$.}
\end{deluxetable}

\begin{deluxetable}{lc}
\tabletypesize{\scriptsize} 
\tablecaption{Subsample bootstrap run parameters\label{tbl-scode}} 
\tablewidth{0pt} 
\tablehead{
\colhead{Run code} & \colhead{$f_s$} } 
\startdata 
S1 & 0.125 \\
S2 & 0.25 \\ 
S3 & 0.5 \\
S4 & 0.75 \\
\enddata
\tablecomments{$f_s$ is the fraction of data deleted in each subsample realization}
\end{deluxetable}

\begin{deluxetable}{lcc}
\tabletypesize{\scriptsize}
\tablecaption{Bootstrap performance statistics\label{tbl-perform}}
\tablewidth{0pt}
\tablehead{
\colhead{Code} & \colhead{$v_f$} & \colhead{$v_{MSE}$} \\
\colhead{} & \colhead{} & \colhead{($\times 10^{-10}$)}
}
\startdata
M1 & 1.00 & 2.95 \\
M2 & 1.01 & 2.28 \\
M3 & 1.00 & 2.19 \\
M4 & 1.02 & 2.36 \\
S1 & 0.157 & 6.43 \\
S2 & 0.319 & 5.15 \\
S3 & 0.782 & 3.42 \\
S4 & 2.03  & 3.04 \\
\enddata

\tablecomments{The bootstrap performance measures, $v_f$ and
$v_{MSE}$, are defined in the main text.}
\end{deluxetable}

\begin{deluxetable}{lcc}
\tabletypesize{\scriptsize}
\tablecaption{MSE of CDF fit to parent distribution\label{tbl-cdf}}
\tablewidth{0pt}
\tablehead{
\colhead{Code} & \colhead{$c_{MSE}$} \\
\colhead{} & \colhead{($\times 10^{-4}$)}
}
\startdata
M1 & 11.1 \\
M2 & 25.5 \\
M3 & 5.35 \\
M4 & 8.07 \\
\enddata

\tablecomments{The MSE of the CDF fit to the parent distribution,
$c_{MSE}$, is defined in the main text. The value tabulated here is
for visibility row 1000 in cross-correlation $LR$. For reference, the
corresponding value from the Monte Carlo sample is $3.48 \times
10^{-4}$.}
\end{deluxetable}



\clearpage

\clearpage

\clearpage

\clearpage

\clearpage

\clearpage

\clearpage

\clearpage

\clearpage

\clearpage

\clearpage

\clearpage

\clearpage

\begin{thebibliography}{}

\bibitem[Perley, Schwab, \& Bridle(1989)] {per89} Perley, R.A.,
Schwab, F.R., \& Bridle, A.H., editors 1989, ASP Conf. Ser. 6,
Synthesis Imaging in Radio Astronomy, A Collection of Lectures from
the Third NRAO Synthesis Imaging Summer School, ed. by R.A. Perley,
F.R. Schwab, \& A.H. Bridle (San Francisco: ASP)

\bibitem[Cawthorne et al.(1993)] {caw93} Cawthorne, T.V., Wardle,
J.F.C., Roberts, D.H., \& Gabuzda, D.C. 1993, \apj, 416, 519

\bibitem[Chernick(1999)] {che99} Chernick, M.R. 1999, Bootstrap
Methods: A Practitioner's Guide (New York: Wiley)

\bibitem[Conway \& Kronberg(1969)] {con69} Conway, R.G., \& Kronberg,
P.P. 1969, \mnras, 142, 11

\bibitem[Cornwell(1995a)] {cor95a} Cornwell, T.J. 1995a, AIPS++ Note
183, http://aips2.nrao.edu/docs/notes/183/183.html

\bibitem[Cornwell(1995b)] {cor95b} Cornwell, T.J. 1995b, AIPS++ Note
184, http://aips2.nrao.edu/docs/notes/184/184.html

\bibitem[Cornwell \& Wieringa(1996)] {cor96} Cornwell, T.J., \&
Wieringa, M.H. 1996, AIPS++ Note 189, http://aips2.nrao.edu/docs/notes/189/189.html

\bibitem[Cornwell \& Fomalont(1999)] {cor99} Cornwell, T., \&
Fomalont, E.B. 1999, in ASP Conf. Ser. 180, Synthesis Imaging in Radio
Astronomy II, A Collection of Lectures from the Sixth NRAO/NMIMT
Synthesis Imaging Summer School, ed. by G.B. Taylor, C.L. Carilli, \&
R. A. Perley (San Francisco:ASP), 187

\bibitem[Cotton et al.(1984)] {cot84} Cotton, W.D., Geldzahler, B.J.,
Marcaide, J.M., Shapiro, I.I., Sanroma, M., \& Rius, A. 1984, \apj,
286, 503

\bibitem[Cotton(1993)] {cot93} Cotton, W.D. 1993, \aj, 106, 1241

\bibitem[Davison \& Hinkley(1997)] {dav97} Davison, A.C., \& Hinkley,
D.V. 1997, Bootstrap Methods and their Applications (Cambridge: CUP)

\bibitem[Efron(1979)] {efr79} Efron B., 1979, Annals of Statistics, 7, 1

\bibitem[Esch et al.(2004)] {esc04} Esch, D.N., Connors, A., Karovska,
M., \& van Dyk, D.A. 2004, \apj, 610, 1213

\bibitem[Hamaker, Bregman, \& Sault(1996)] {ham96a} Hamaker, J.P.,
Bregman, J.D., \& Sault, R.J. 1996, A\&AS, 117, 137

\bibitem[Hamaker \& Bregman(1996)] {ham96b} Hamaker, J.P., \& Bregman,
J.D. 1996, A\&AS, 117, 161

\bibitem[Hamaker(2000)] {ham00} Hamaker, J.P. 2000, A\&AS, 143, 515

\bibitem[H\"ogbom(1974)] {hog74} H\"ogbom, J.A. 1974, A\&AS, 15, 417

\bibitem[Kay(1993)] {kay93} Kay, S.M. 1993, Fundamentals of
Statistical Signal Processing: Estimation Theory (New Jersey: Prentice
Hall)

\bibitem[Kellermann \& Thompson(1985)] {kel85} Kellermann, K.I., \&
Thompson, A.R. 1985, Science, 229, 123

\bibitem[Kemball(1993)] {kem93} Kemball, A.J. 1993, Ph.D. thesis,
Rhodes University

\bibitem[Kemball, Diamond, \& Cotton(1995)] {kem95} Kemball, A.J.,
Diamond, P.J., \& Cotton, W.D. 1995, A\&AS, 110, 383

\bibitem[Kemball, Diamond, \& Pauliny-Toth(1996)] {kem96} Kemball,
A.J., Diamond, P.J., \& Pauliny-Toth, I.I.K. 1996, \apj, 464, L55

\bibitem[Kemball and Diamond(1997)] {kem97} Kemball, A.J., \& Diamond,
P.J. 1997, \apj, 481, L111

\bibitem[Kemball(1999)] {kem99} Kemball, A.J. 1999, in ASP
Conf. Ser. 180, Synthesis Imaging in Radio Astronomy II, A Collection of
Lectures from the Sixth NRAO/NMIMT Synthesis Imaging Summer School,
ed. by G.B. Taylor, C.L. Carilli, \& R. A. Perley (San Francisco:ASP),
499

\bibitem[Kemball(2002)] {kem02} Kemball, A.J. 2002, in ASP Conf
Ser. 206, Cosmic Masers: From Protostars to Black Holes,
ed. V. Migenes \& M.J. Reid (San Francisco: ASP), 359

\bibitem[Lahiri(2003)] {lah03} Lahiri, S.N. 2003, Resampling Methods
for Dependent Data (New York: Springer-Verlag)

\bibitem[Leppanen, Zensus, \& Diamond(1995)] {lep95} Leppanen, K.J.,
Zensus, J.A., \& Diamond, P.J. 1995, \aj, 110, 2479

\bibitem[Maitra(1997)] {mai97} Maitra, R. 1997, J. Comp. Graph. Stat., 6, 132

\bibitem[Morris, Radhakrishnan, \& Seielstad(1964)] {mor64} Morris,
D., Radhakrishnan, V., \& Seielstad, G.A. 1964, \apj, 139, 551

\bibitem[Noordam(1995)] {noo95} Noordam, J.E. 1995, AIPS++ Note 182,
http://aips2.nrao.edu/docs/notes/182/182.html

\bibitem[Noordam(1996)] {noo96} Noordam, J.E. 1996, AIPS++ Note 185, 
http://aips2.nrao.edu/docs/notes/185/185.html

\bibitem[Politis, Romano, \& Wolf(1999)] {pol99} Politis, D.N.,
Romano, J.P., \& Wolf, M. 1999, Subsampling (New York:
Springer-Verlag)

\bibitem[Readhead \& Wilkinson(1978)] {rea78} Readhead, A.C.S, \&
Wilkinson, P.N. 1978, \apj, 223, 25

\bibitem[Roberts et al.(1984)] {rob84} Roberts, D.H., Potash, R.I.,
Wardle, J.F.C., Rogers, A.E.E., \& Burke, B.F. 1984, in IAU Symp. 110,
VLBI and Compact Radio Sources, ed. R. Fanti, K. Kellermann, \&
G. Setti (Dordrecht:Reidel), 35

\bibitem[Roberts, Brown, \& Wardle(1991)] {rob91} Roberts, D.H.,
Brown, L.F., \& Wardle, J.F.C. 1991, in ASP Conf. Ser. 19, Radio
Interferometry: Theory, Techniques, and Applications, ed. by
T. Cornwell \& R.A. Perley (San Francisco: ASP), 281

\bibitem[Roberts, Wardle, \& Brown(1994)] {rob94} Roberts, D.H.,
Wardle, J.F.C, \& Brown, L.F. 1994, \apj, 427, 718

\bibitem[Rogers et al.(1974)] {rog74} Rogers, A.E.E, Hinteregger,
H.F., Whitney, A.R., Counselman, C.C., Shapiro, I.I., Wittels, J.J.,
Klemperer, W.K., Warnock, W.W., Clark, T.A., \& Hutton, L.K. 1974,
\apj, 193, 293

\bibitem[Sault, Hamaker, \& Bregman(1996)] {sau96} Sault, R.J.,
Hamaker, J.P., \& Bregman, J.D. 1996, A\&AS, 117, 149

\bibitem[Shao \& Tu(1995)] {sha95} Shao, J., \& Tu, D. 1995, The
Jackknife and Bootstrap (New York: Springer-Verlag)

\bibitem[Singh(1981)] {sin81} Singh, K. 1981, Annals of Statistics, 9, 1187

\bibitem[Taylor, Carilli, \& Perley(1999)] {tay99} Taylor, G.B.,
Carilli, C.L., \& Perley, R.A., editors 1999, ASP Conf. Ser. 180, Synthesis
Imaging in Radio Astronomy II, A Collection of Lectures from the Sixth
NRAO/NMIMT Synthesis Imaging Summer School, ed. by G.B. Taylor,
C.L. Carilli, \& R. A. Perley (San Francisco:ASP)

\bibitem[Thompson, Moran, \& Swenson(2001)] {tho01} Thompson, A.R.,
Moran, J.M., \& Swenson Jr., G.W. 2001, Interferometry and Synthesis
in Radio Astronomy (2d ed.; New York: Wiley)

\bibitem[Wardle \& Roberts(1994)] {war94} Wardle, J.F.C., \& Roberts,
D.H. 1994, in Compact Radio Sources, ed. by J.A. Zensus \&
K.I. Kellermann (Green Bank: NRAO), 217

\bibitem[Weiler(1973)] {wei73} Weiler, K.W. 1973, AA, 26, 403

\end{thebibliography}
\end{document}